\documentclass[fleqn,useAMS,usenatbib]{mnras}

\usepackage[T1]{fontenc}
\usepackage{ae,aecompl}

\usepackage{pdflscape}
\usepackage{graphicx}	
\usepackage{amsmath}	
\usepackage{amssymb}	
\usepackage{color}

\newcommand{\kms}{\hbox{${\rm km\;s}^{-1}$}}

\newcommand{\hi}{H~\textsc{i}}   

\newcommand{\vradio}{\ensuremath{V_{\mathrm{radio}}}}
\newcommand{\vopt}{\ensuremath{V_{\mathrm{optical}}}}

\newcommand{\amax}{\ensuremath{a_{\epsilon}}}

\newcommand{\avis}{\ensuremath{a_{\mathrm{vis}}}}
\newcommand{\atwomax}{\ensuremath{A_{2,{\rm max}}}}

\newcommand{\Msun}{\ensuremath{M_{\sun}}}
\newcommand{\Mstar}{\ensuremath{M_{\star}}}
\newcommand{\logmstar}{\ensuremath{\log \, (M_{\star}/M_{\sun})}}

\newcommand{\MHI}{\ensuremath{M_{\mathrm{H} \textsc{i}}}}

\newcommand{\mhi}{\ensuremath{m_{21,\mathrm{c}}}}
\newcommand{\fgas}{\ensuremath{f_{\mathrm{gas}}}}
\newcommand{\logfgas}{\ensuremath{\log f_{\mathrm{gas}}}}
\newcommand{\btc}{\ensuremath{B_{\mathrm{tc}}}}
\newcommand{\bmv}{\ensuremath{B\! - \!V}}
\newcommand{\bmvtc}{\ensuremath{(B\! - \!V)_{\mathrm{tc}}}}

\newcommand{\gmr}{\ensuremath{g\! - \!r}}

\newcommand{\sfourg}{S\ensuremath{^{4}}G}
\newcommand{\fbar}{\ensuremath{f_{\mathrm{bar}}}}
\newcommand{\fSB}{\ensuremath{f_{\mathrm{SB}}}}
\newcommand{\fSAB}{\ensuremath{f_{\mathrm{SAB}}}}

\title[Frequency of Bars]{The Dependence of Bar Frequency 
on Galaxy Mass, Colour, and Gas Content -- and Angular Resolution -- in the Local Universe}
\author[P. Erwin]{Peter Erwin$^{1,2}$\thanks{E-mail: erwin@mpe.mpg.de} \\
$^{1}$Max-Planck-Insitut f\"{u}r extraterrestrische Physik, Giessenbachstrasse, 85748 Garching, Germany \\
$^{2}$Universit\"{a}ts-Sternwarte M\"{u}nchen, Scheinerstrasse 1, D-81679 M\"{u}nchen, Germany}

\date{Accepted XXX. Received YYY; in original form ZZZ}

\pubyear{2018}

\begin{document}
\label{firstpage}
\pagerange{\pageref{firstpage}--\pageref{lastpage}}
\maketitle

\begin{abstract} 

I use distance- and mass-limited subsamples of the \textit{Spitzer}
Survey of Stellar Structure in Galaxies (\sfourg) to investigate how the
presence of bars in spiral galaxies depends on mass, colour, and gas
content and whether large, SDSS-based investigations of bar frequencies
agree with local data.  Bar frequency reaches a maximum of $\fbar
\approx 0.70$ at $\Mstar \sim 10^{9.7} \Msun$, declining to both lower
and higher masses. It is roughly constant over a wide range of colours
($\gmr \approx 0.1$--0.8) and atomic gas fractions ($\log (\MHI /
\Mstar) \approx -2.5$ to 1). Bars are thus as common in blue, gas-rich
galaxies are they are in red, gas-poor galaxies.  This is in sharp
contrast to many SDSS-based studies of $z \sim 0.01$--0.1 galaxies,
which report \fbar{} increasing strongly to higher masses (from
$M_{\star} \sim 10^{10}$ to $10^{11} \Msun$), redder colours, and lower
gas fractions. The contradiction can be explained if SDSS-based studies
preferentially miss bars in, and underestimate the bar fraction
for, lower-mass (bluer, gas-rich) galaxies due to poor spatial
resolution and the correlation between bar size and stellar mass.
Simulations of SDSS-style observations using the \sfourg{} galaxies as a
parent sample, and assuming that bars below a threshold angular size of
twice the PSF FWHM cannot be identified, successfully reproduce typical
SDSS \fbar{} trends for stellar mass and gas mass ratio. Similar
considerations may affect high-redshift studies, especially if bars
grow in length over cosmic time; simulations 
suggest that high-redshift bar fractions may thus be systematically
underestimated.

\end{abstract}


\section{Introduction}\label{sec:intro} 

Stellar bars are an important -- and often visually striking --
component of many disc galaxies. They are believed to be key drivers of
secular evolution, in part because they can act as both sources and
sinks for angular momentum, redistributing stars, gas, and dark matter
within galaxies \citep[
e.g.,][]{athanassoula02b,athanassoula03,weinberg02,kormendy04,
holley-bockelmann05,debattista06,sellwood14}. Perhaps the most visible aspect
of this is bar-driven gas inflow and subsequent star formation in the
central regions of galaxies, which has been suggested to play a key role
in building up bulges and pseudobulges. Bar-driven gas inflow has also
long been implicated in the fuelling of AGN \citep[although confirmation
of this last idea has proved somewhat elusive;
e.g.,][]{shlosman89,shlosman90,ho97,laine02,lee12b,cheung15,goulding17}.

In addition to their direct effects on their host galaxies, the mere
presence or absence of bars can provide important insights into galactic
evolution. The discovery from early simulations that $N$-body discs
routinely become bar-unstable raised the question of why some disc
galaxies did \textit{not} have bars. The demonstration that massive
halos could stabilize discs against bar formation
\citep[e.g.,][]{ostriker73,hohl76} was one reason for the growing
acceptance of dark-matter halos in the 1970s, and bar-instability
criteria, especially based on \citet{efstathiou82}, have played a major
role in semi-analytic models of galaxy evolution \citep[e.g.,][]{mo98}.
Ironically, subsequent numerical experiments with live dark-matter halos
have shown that halos can actually \textit{amplify} bar growth, since
halo particles can absorb angular momentum from bar instabilities and
help them grow
\citep[e.g.,][]{athanassoula02b,athanassoula08,saha13,sellwood16}.
Nevertheless, the fact that instabilities in sufficiently cool discs
readily give rise to bars means that the appearance of bars in galaxies
at high redshifts can be a useful -- and easily visible -- indicator of
when and how rapidly galaxy discs become kinematically cool
\citep[e.g.,][]{sheth12}.

Bars are thus important both for how they can reshape galaxies and as
general diagnostics for the evolution of discs. But in order to have a clear
understanding of how important bars really are for galaxy evolution in
general, we need to know what fraction of discs really do (or do not)
have bars. Ideally, we also want to know what \textit{kinds} of discs
are more (or less) likely to host bars.

In the local universe (roughly, closer than $\sim 50$ Mpc), the fraction
of disc galaxies with stellar bars has been reported to be roughly
$2/3$ (or $1/3$ if only so-called ``strong'' bars are counted). A nice
overview of past studies can be found in the Introduction of
\citet{sheth08}; they write, ``We conclude that the consensus value of
the local barred fraction is $\fbar \sim 0.3$ for strongly barred
systems, and $\fbar \sim 0.65$ for all barred galaxies, and that these
values are so well known that they have not changed significantly in
over four decades.''

The advent of multi-band imaging (and accompanying spectroscopy) from
the Sloan Digital Sky Survey \citep[SDSS;][]{york00} has opened up
impressively large new fields of analysis for the moderately nearby
universe (i.e., beyond $\sim 50$ Mpc and out to $z \sim 0.1$). Studies
using increasingly large numbers of galaxies have looked at the question
of how common bars are, beginning with \citet{barazza08} and
\citet{aguerri09}, who analyzed samples of 1144 and 3060 low-inclination
discs, respectively, and then \citet{nair10a,nair10b}, using
approximately 7700 galaxies (out of a larger sample of 14,700 galaxies).
The largest samples have come from successors to the Galaxy Zoo project
\citep{lintott08}, in particular Galaxy Zoo 2 \citep[hereafter
GZ2][]{willett13}, with samples of between 13,000 and 15,000
low-inclination discs and bar classifications via an innovative
citizen-science approach \citep[e.g.,][]{masters11,skibba12,cheung13}.
These studies have generally found rather low bar fractions: most often
$\sim 25$--30\%, though values as high as $\sim 50$\% have come from
some surveys \citep[e.g.,][]{barazza08,aguerri09,yoshino15}. Several of
the SDSS-based studies have cautioned that their results are perhaps
applicable primarily to ``strong'' bars, however those might be defined.
The fact that the local strong-bar (SB) frequency is $\sim 30$\% makes this
an appealing argument, since it more or less matches the most common
SDSS-based bar frequencies.

Many of these studies have also investigated what kinds of disc galaxies
are more -- or less -- likely to host bars. This has yielded some rather
striking results: bars are apparently much more common in red, massive,
and gas-poor galaxies than they are in blue, low-mass, and gas-rich
galaxies
\citep[e.g.,][]{masters11,masters12,skibba12,oh12,lee12a,cheung13,
gavazzi15,consolandi16b,cervantes-sodi17}. Only a few SDSS-based studies
have found different -- or more complex -- trends
\citep[e.g.,][]{barazza08,nair10b,mendez-abreu12}.

Studies using \textit{HST} imaging
\citep[e.g.,][]{abraham99,sheth03,jogee04,elmegreen04,sheth08,cameron10,
sheth12,melvin14,simmons14} have looked at how common bars are at higher
redshifts (mostly $z \sim 0.4$--1.0), to see if we can observe evolution
in the bar fraction. Various methodologies for bar detection have been
used, including a version of the GZ2 citizen-science approach
\citep{melvin14,simmons14}. Most of these studies have reported a clear
\textit{decrease} in the bar fraction toward higher redshifts. At least
some have found evidence that at any given redshift there are more bars
in higher-mass galaxies than in lower-mass galaxies, and also that the
bar fraction reaches present-day levels first for higher-mass samples
\citep{sheth08,melvin14}. This seems consistent with the strong
mass-dependence of bar frequency seen in the SDSS studies, as if the bar
fraction saturated for high-mass galaxies some time ago, and is still
low (and increasing?) for lower-mass galaxies.

The \textit{Spitzer} Survey of Stellar Structure in Galaxies
\citep[\sfourg;][]{sheth10} has made available high-S/N near-IR images
for over 2000 local galaxies. Approximately 1300 of these are disc
galaxies with inclinations low enough for reliable detection and
analysis of bars. \citet{dg16a,dg16b} have recently presented a detailed
analysis of bars identified by \citet{buta15} and measured by
\citet{herrera-endoqui15}. Overall, the local bar fraction is apparently
still in the 60--70\% range; \citet{buta15} report a bar
fraction of ``about 2/3''. The upper panel of Fig.~19 in \citet{dg16a}
shows how bar frequency in the entire face-on sample depends on stellar
mass: there is a general increase in \fbar{} from the lowest stellar
masses to a broad peak or plateau with $\fbar \sim 0.6$, extending from
$\logmstar \sim 9.2$--10.6, with a weak maximum at $\sim 9.7$. This
disagrees with most of the SDSS-based studies. In particular, the strong
increase in \fbar{} from $\logmstar \sim 10$ to 11 seen for most
large SDSS samples \citep{nair10b,masters12,oh12,skibba12,melvin14} is
\textit{not} seen in the local sample, nor is the strong increase in
\fbar{} to lower stellar masses reported by \citet{nair10b}.

We have, therefore, several related puzzles. Can we resolve the apparent dichotomy
in overall bar frequency between local samples ($\fbar \sim 0.65$) and 
larger, more distant SDSS-based samples ($\fbar \sim 0.25$--0.5)? Is the difference
just due to the the SDSS studies only detecting strong bars? Do the apparently
strong correlations between bar frequency and stellar mass, colours, and gas richness
seen in the SDSS studies also exist for local galaxies? If not, why not? And
what, if anything, does this imply for studies of bars at high redshifts?

In this paper, I investigate these questions by seeing whether \sfourg{}
galaxies -- restricted to volume- and mass-limited subsamples and with
added data from the literature (Section~\ref{sec:data}) -- show the same
bar-frequency trends with stellar mass, colour, and gas richness that the
large SDSS samples do. Since the answer to this question turns out to be
a rather strong ``No'' (Section~\ref{sec:fbar}), I focus on whether the
disagreement could be due to problems of spatial resolution, given that
the average galaxy in an SDSS-based survey is almost ten times further
away than the average \sfourg{} galaxy. This requires taking a closer
look at bar \textit{sizes}, which prove \citep[as previously noted
by][]{dg16a} to have a strong dependence on stellar mass
(Section~\ref{sec:bar-sizes}). I then argue that a dependence of bar
size on stellar mass can straightforwardly translate into a
mass-dependent, differential bar-detection bias, and that the
combination of the actual sizes of bars and the typical resolution of
SDSS images make this potential bias a very real possibility, one which
could explain the typical dependence of bar frequency on stellar mass
seen in SDSS-based studies. Since colour and gas richness (and certain
other properties such as specific star-formation rate) have strong
correlations with stellar mass, this can also explain, at least partly,
the reported dependence of bar frequency on colour, gas mass ratio, and
star-formation rates. By treating the \sfourg{} galaxies as parent
sample and generating bootstrapped mock samples of galaxies observed at
typical SDSS distances, I show (Section~\ref{sec:sims}) that the
reported SDSS-based bar-fraction trends for stellar mass and gas mass
ratio can be qualitatively -- and least partly quantitatively --
generated just by assuming that bars with projected angular sizes less
than twice the FWHM of the PSF go undetected. I consider some additional
issues -- in particular, the implications for high-redshift studies of
bar frequencies -- in Section~\ref{sec:discuss}.

Where necessary, I assume a standard $\Lambda$CDM cosmology based on
\citet{bennett13}, with $H_{0} = 70$ \kms{} Mpc$^{-1}$, $\Omega_{m} = 0.29$,
and $\Omega_{\Lambda} = 0.71$. 

To aid in reproducibility, there is a Github repository containing data
files, code, and Jupyter notebooks for creating the figures, fits, and
simulations used in this paper. The repository is available at
\url{https://github.com/perwin/s4g_barfractions} and also at \url{https://doi.org/10.5281/zenodo.804909}.

\section{Data and Sample Definitions for Local Galaxies}\label{sec:data} 

\subsection{Parent Sample and Data Sources} 
\label{sec:data-sources}

The best local sample for assessing bar frequencies and correlations is
undoubtedly the \textit{Spitzer} Survey of Stellar Structure in Galaxies
\citep[\sfourg;][]{sheth10}, both because of its size and because of its
use of near-infrared imaging, which minimizes the possibility of missing
bars due to the confusion introduced by dust extinction and star
formation. The entire sample has been subjected to a consistent
classical morphological analysis -- including bar classifications -- by
\citet{buta15}, with extensive quantitative bar measurements and
analyses by \citet{herrera-endoqui15} and \citet{dg16a,dg16b}. It does suffer
from being a magnitude- and diameter-limited sample (e.g.,
Figure~\ref{fig:mstar-r25-vs-distance}), which prevents it from being
fully volume-limited; however, the relatively faint magnitude limit ($B
\leq 15.5$) ensures that reasonable volume- and mass-limited subsamples
can still be derived from it; the details of this are discussed in
Section~\ref{sec:volume-mass-limits}.

I start with the \sfourg{} subsample of non-edge-on disc galaxies
defined by \citet{dg16a}. This takes the version of \sfourg{} classified
by \citet{buta15} and then eliminates elliptical galaxies and disc
galaxies with inclinations $> 65\degr$, leaving a total of 1344
galaxies. Distances and stellar masses for these galaxies are taken from
\citet{munoz-mateos15}. The distances are based on redshift-independent
NED distances for 79\% of the total sample, and on Hubble-flow distances
for the rest (assuming $H_{0} = 71$). Ten galaxies missing distances and
stellar masses in \citet{munoz-mateos15} were then removed, along with
twelve more which had very uncertain distances (i.e., redshifts $< 500$
\kms{} and no alternate distance measurements) or optical diameters
smaller than the original \sfourg{} limit ($D_{25} = 1.0\arcmin$),
leaving a total of 1322 galaxies. This is the Parent Disc Sample.

The \sfourg{} sample as studied in the literature suffers from
incompleteness in terms of Hubble types, stemming from the fact that it
was defined to have a redshift limit of 3000 \kms{} using \textit{radio}
radial velocities. Galaxies without such velocities were thus excluded,
even if they had \textit{optical} radial velocities that would put them
in the sample. This limitation creates a morphological bias: gas-poor
early-type galaxies, including both ellipticals and S0s, are much more
likely to be missing than are spiral galaxies.\footnote{\citet{buta15}
noted that improved observing efficiency had allowed the addition of 21
gas-poor galaxies to the original sample.} Using the February 2017
version of HyperLeda and the basic \sfourg{} sample definition of
\citet{sheth10}, there are 167 S0 galaxies with $\vradio < 3000$ \kms{}
(156 of which are in the complete \sfourg{} sample\footnote{The missing
11 S0s are presumably galaxies whose radio velocities were added to the
HyperLeda database after the original \sfourg{} definition in late 2007;
see \citet{sheth10}.}) and 315 without $\vradio$ values but with $\vopt
< 3000$ \kms; this contrasts with 2350 spirals and irregulars with
$\vradio < 3000$ \kms{} and only 273 without $\vradio$ values. So there
is an incompleteness of $\approx 67$\% for S0 galaxies versus only 10\%
for later types. (Further \textit{Spitzer} observations to fill in the
missing ellipticals and S0s have been obtained by the \sfourg{} team --
e.g., \citealt{sheth13}, \citealt{knapen14} -- but the corresponding
morphological analyses and measurements are not yet available, so I
confine myself to the original, spiral-dominated sample.)

Since there is some evidence for differences in bars in S0s versus bars
in spirals, both in terms of frequencies and also in terms of bar
strengths \citep[e.g.,][]{aguerri09,buta10,dg16a}, it makes sense to
separate the highly incomplete S0 subsample from the mostly complete
spirals. Consequently, I exclude S0 galaxies from the main samples for
analysis in this paper, by retaining only galaxies with optical Hubble
type $T > -0.5$. The result is the Parent Spiral Sample, with 1220
galaxies (note that this name is slightly misleading, since the sample
includes $\sim 150$ irregular galaxies). 

Whether a galaxy is considered barred or not comes from Table~2 of
\citet{herrera-endoqui15}: if a bar and its accompanying measurements
are among the features listed for that galaxy, then it is deemed to be
barred. This is in turn based on the visual bar classifications of
\citet{buta15}, except that 117 of the latter galaxies with barred
classifications do not have barred measurements in
\citealt{herrera-endoqui15}; these are primarily galaxies with the very
weakest \citealt{buta15} subclassification (S\textunderscore{A}B). Since
considerations of bar \textit{size} turns out to be important for
understanding reported bar frequencies, I stick with the measurements of
\citealt{herrera-endoqui15} and treat the ``missing'' \citealt{buta15}
barred galaxies as unbarred; the bar frequencies I report for \sfourg{}
subsamples may thus be slight \textit{underestimates}.

As a simple measure of bar ``strength'' I use the basic
subclassifications of bars into strong (SB) and weak (SAB) categories as
listed by \citet{buta15}. Measurements of bar sizes are taken from
\citet{herrera-endoqui15}. The latter authors provided both ``visual''
(\avis) and maximum-ellipticity-based (\amax) measurements of bar
lengths (semi-major axes).  \citet{dg16a} showed that these measurements
are generally consistent with each other (e.g., their Fig.~9), with
\amax{} on average marginally smaller than \avis. Since
maximum-ellipticity lengths are not available for $\sim 22$\% of the
bars (usually cases where low S/N and/or strong star formation made
ellipse fits unreliable), I use the \avis{} values. To compute
deprojected sizes, I use bar position angles from
\citet{herrera-endoqui15} and disc orientations from \citet{salo15}.

For galaxy colours, I take \bmvtc{} values from HyperLeda; these are
whole-galaxy colours, corrected for Galactic and internal extinction.
These are then converted into \gmr{} colours, since that is the most
common colour used by published SDSS studies of bar fractions. The colour
conversion -- and incompleteness corrections necessitated by the fact
that about half the galaxies do not have \bmvtc{} colours in HyperLeda --
are discussed in more detail in Appendix~\ref{sec:colors}.

Finally, I also take \hi{} fluxes from HyperLeda in order to determine
\hi{} masses and the gas mass ratio $\fgas = \MHI / \Mstar$. The \hi{}
mass is calculated from the HyperLeda \mhi{} values using the standard
relation of
\citet{giovanelli88}:
\begin{equation} 
\MHI = 2.356 \times 10^{5} D^{2} 10^{0.4 (17.40 - \mhi)}, 
\end{equation} 
where $D$ is the distance in Mpc and \mhi{} is the corrected \hi{} magnitude from
HyperLeda.
In order to compare bar frequencies as a function of gas mass ratio with
the results of \citet{masters12}, I define \fgas{} as the ratio of
atomic hydrogen to stellar mass, \textit{without} any corrections for
the presence of helium or metals. HyperLeda \mhi{} values exist for
approximately 97\% of the \sfourg{} galaxies in the Parent Disc Sample
(1290 out of 1334 galaxies), so incompleteness is not a meaningful issue.
The distributions of \gmr{} and \fgas{} values as a function of stellar
mass in the Parent Spiral Sample are shown in Figure~\ref{fig:gmr-and-fgas-vs-mstar}.

\subsection{Correcting for Incompleteness: Volume and Mass Limits and Final Sample Definitions}\label{sec:volume-mass-limits} 
 
\begin{figure*}
\begin{center}
\hspace*{-1mm}\includegraphics[scale=0.87]{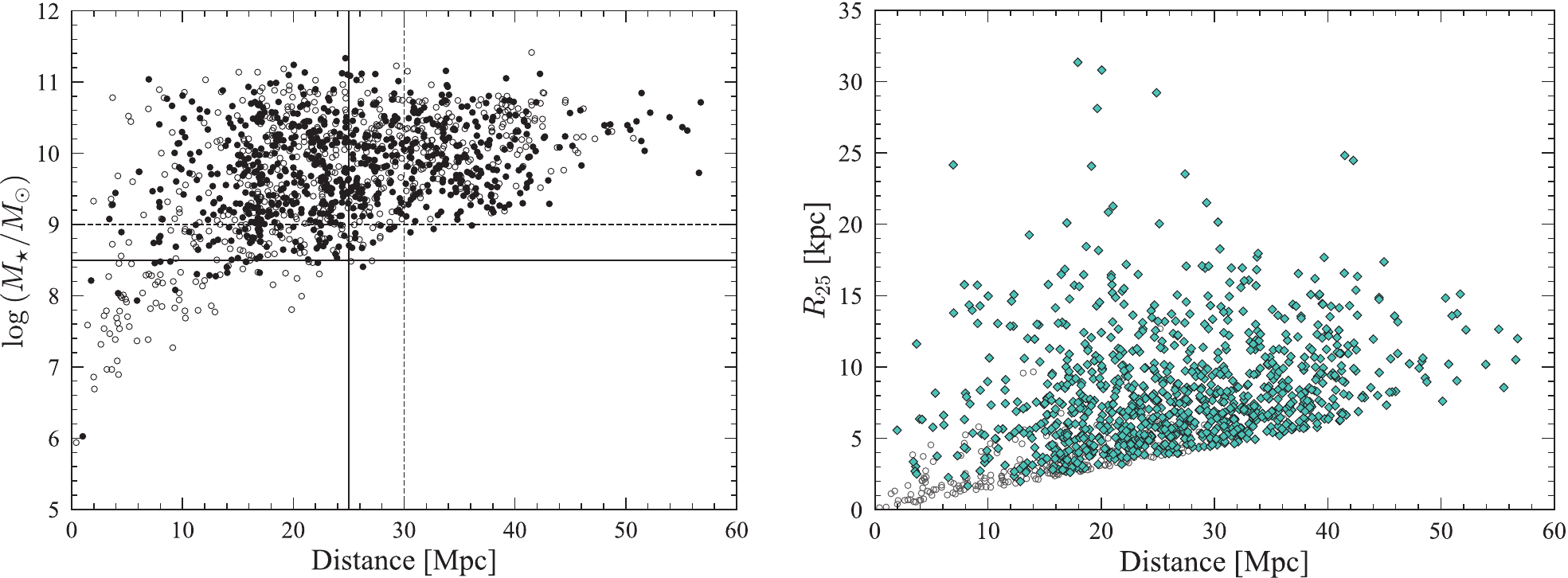}
\end{center}

\caption{Effects of magnitude and angular-size limits on stellar mass
and absolute disc size for the \sfourg{} galaxies in the Parent Disc Sample, based on the
low-inclination \sfourg{} disc sample of \citet{dg16a}. (Five
galaxies with $D > 60$ Mpc are not shown.) Left: Stellar mass versus
distance. Hollow points are unbarred galaxies, filled points are barred.
Lines show distance and stellar-mass cuts used to define the main
volume-limited subsamples for this paper (see Table~\ref{tab:samples}).
Right: $B$-band optical disc radius $R_{25}$ versus distance (spiral
galaxies only). Cyan diamonds indicate galaxies with stellar masses
$\logmstar \geq 9$.\label{fig:mstar-r25-vs-distance}}

\end{figure*}

\begin{figure*}
\begin{center}
\hspace*{-1.5mm}\includegraphics[scale=0.86]{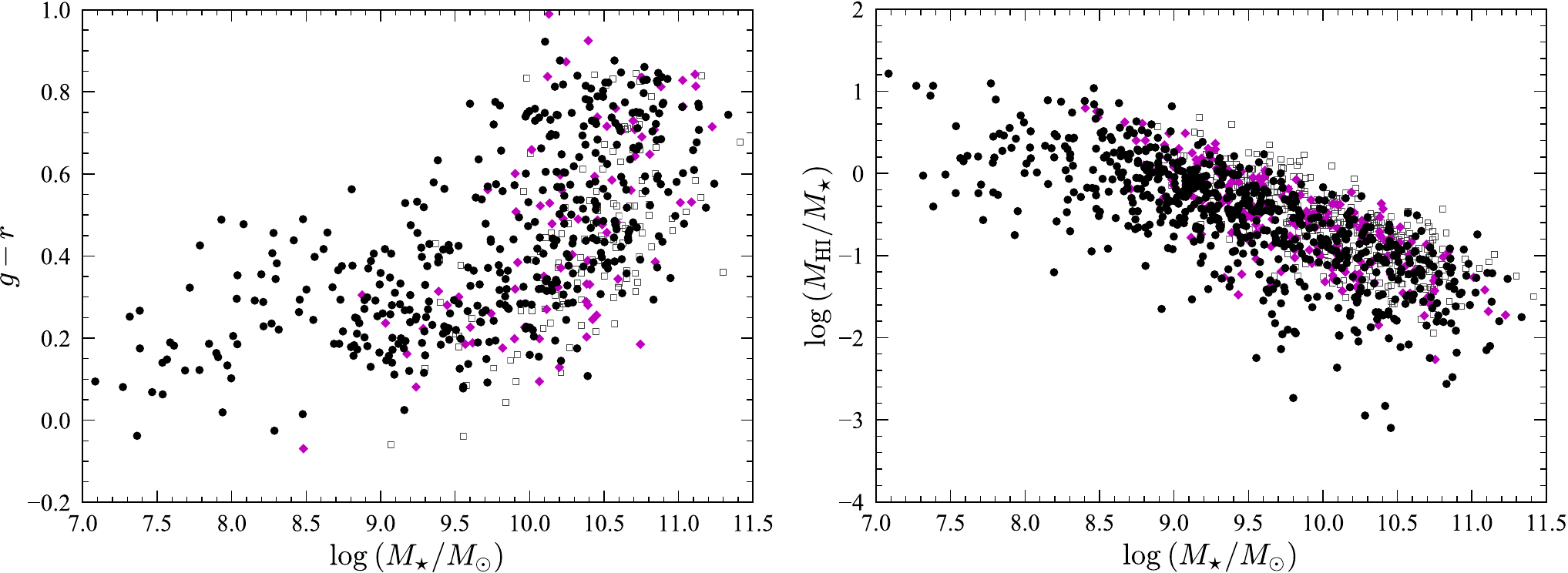}
\end{center}

\caption{Galaxy \gmr{} colours (left) and gas mass fraction \fgas{}
(right) for \sfourg{} spiral galaxies. Galaxies with distances $\leq 25$
Mpc are plotted with filled black circles; galaxies with distances between 25
and 30 Mpc are plotted with filled magenta diamonds, while galaxies at larger
distances (not analyzed in this paper) are plotted with open squares.
\label{fig:gmr-and-fgas-vs-mstar}}

\end{figure*}

The original \sfourg{} sample is magnitude- and diameter-limited
\citep{sheth10}. The magnitude limit ($\btc \leq 15.5$) means that as
one moves out in distance, low-luminosity -- and thus low-mass --
objects drop out of the sample before higher-luminosity (higher-mass)
objects do. This is demonstrated in
Figure~\ref{fig:mstar-r25-vs-distance}, which plots stellar mass against
distance for each galaxy in the Parent Disc Sample. Distance and
stellar-mass limits are therefore needed in order to define complete,
volume-limited subsamples. 

\begin{figure}
\begin{center}
\hspace*{-5mm}\includegraphics[scale=0.48]{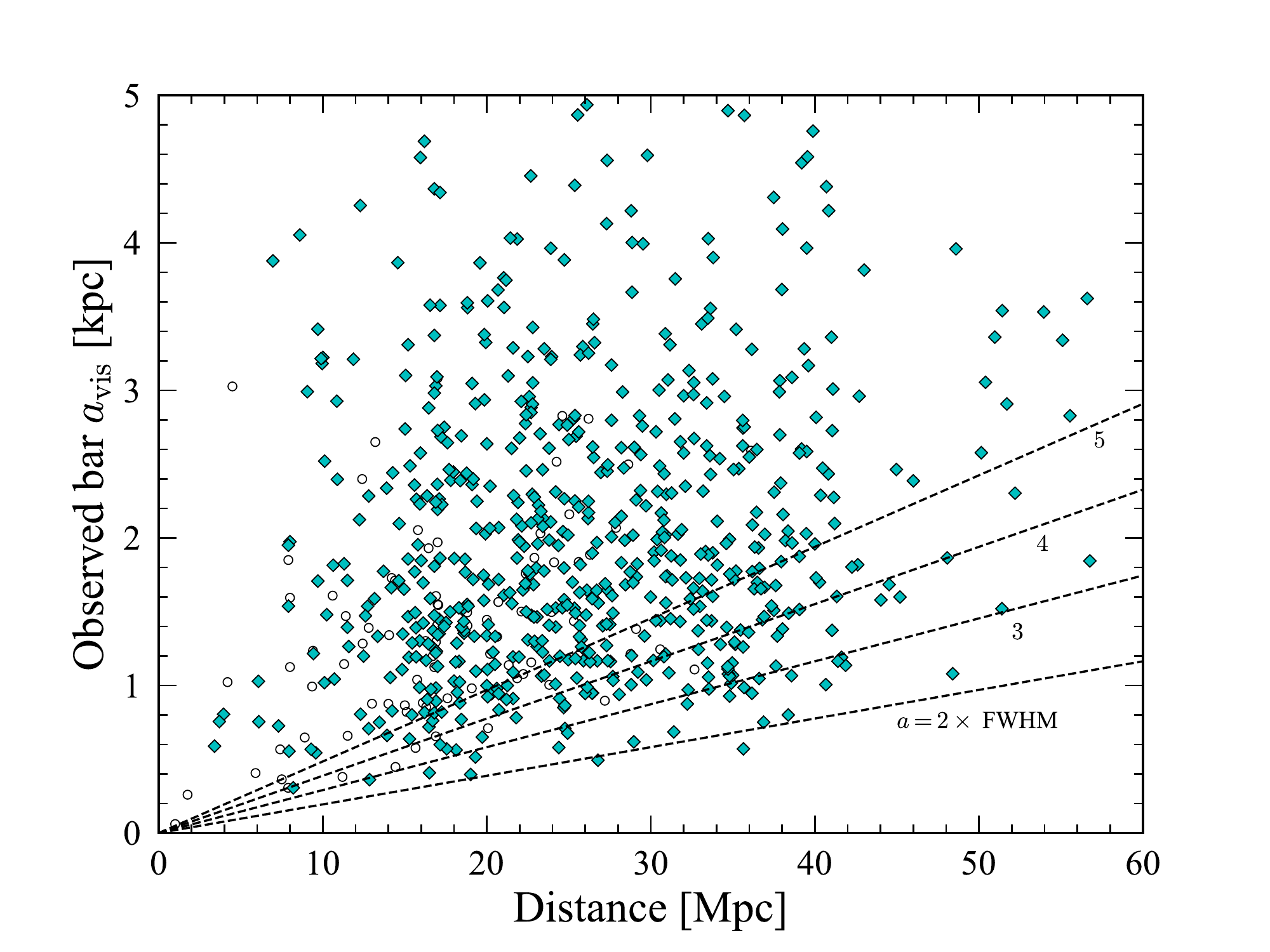}
\end{center}

\caption{Observed bar sizes (semi-major axes) in Parent Disc Sample
(i.e., including S0 galaxies) versus galaxy distance. Dashed lines show
possible resolution limits in multiples of the \sfourg{} PSF's FWHM
($\approx 2\arcsec$). Cyan diamonds are galaxies with $\logmstar \ge
9$.\label{fig:barsize-vs-distance}}

\end{figure}

There is an additional reason to apply a more stringent distance limit
than the $\sim 40$ Mpc cutoff implicit in the original \sfourg{} sample
(defined via a redshift limit of 3000 \kms): bars are harder to identify
in more distant galaxies. Figure~\ref{fig:barsize-vs-distance} shows
observed bar sizes in the Parent Disc Sample, along with lines
indicating possible resolution limits as multiples of the \sfourg{} PSF
\citep[assuming FWHM $\approx 2\arcsec$;][]{salo15}. As previously shown by
\citet{menendez-delmestre07} for a study of bars in local galaxies using
2MASS images, physically smaller bars are less likely to be detected at
larger distances, because their smaller angular size -- combined with
the isophotal-rounding effects of PSF convolution -- makes them harder
to discern. 

\begin{figure}
\begin{center}
\hspace*{-5mm}\includegraphics[scale=0.47]{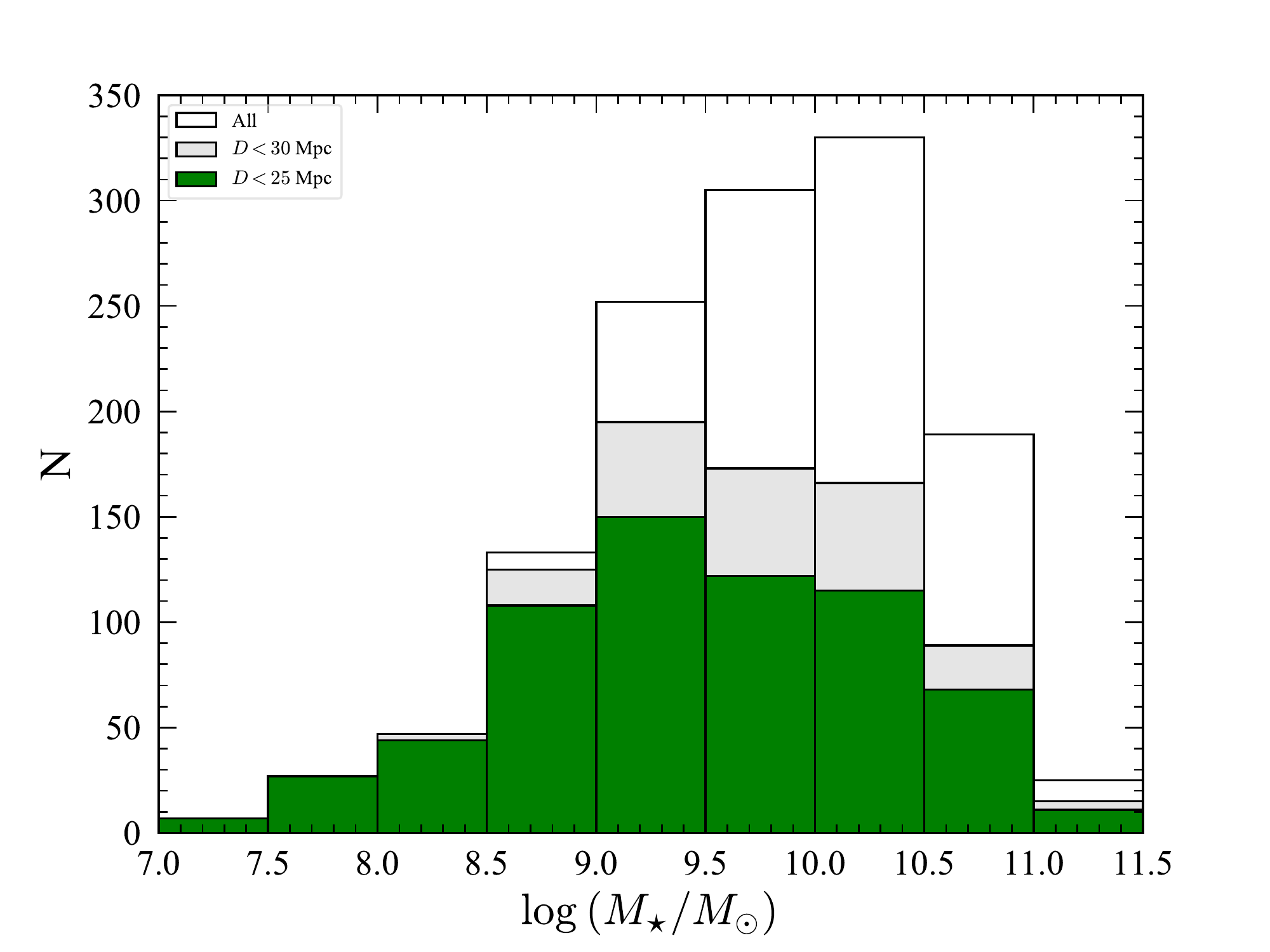}
\end{center}

\caption{Histograms of stellar mass for different distance-limited
subsamples of the \sfourg{} Parent Spiral
Sample.\label{fig:mstar-histogram}}

\end{figure}

Figure~\ref{fig:barsize-vs-distance} suggests that the (visual)
bar-detection criteria used by \citet{buta15} for the \sfourg{} sample
may start to miss smaller bars for distances $\ga 25$--30 Mpc.
Accordingly, I adopt two sets of mass- and distance-limited
subsamples. For plots and analyses of bar frequency and size as a
function of stellar mass -- where the potential incompleteness at the
low-mass ends can be visually excluded by the reader -- I use
distance-limited samples: ``Sample~1'' assumes a limit of 25
Mpc, while ``Sample 2'' uses 30 Mpc. I also construct
mass-limited subsamples: Sample~1m is Sample~1 with the addition of a
stellar-mass limit of $\logmstar \geq 8.5$; Sample~2m is Sample~2 with a
higher stellar-mass cutoff of $\logmstar \geq 9$. These different subsamples are
summarized in Table~\ref{tab:samples}, along with the number of galaxies
in each. The sample definitions are also indicated by the solid
(Samples~1 and 1m) and dashed (Samples~2 and 2m) lines in
Figure~\ref{fig:mstar-r25-vs-distance}. Figure~\ref{fig:mstar-histogram}
shows histograms of stellar mass for the Parent Spiral Sample, Sample~1,
and Sample~2.

Finally, the diameter limit used to define the original \sfourg{} sample
($D_{25} \geq 1.0\arcmin$) means that there is a distance-based bias against \textit{more compact}
galaxies even after mass limits are applied. The right-hand panel of
Figure~\ref{fig:mstar-r25-vs-distance} shows $R_{25}$ in kpc versus
galaxy distance for the Parent Spiral Sample. Because this diameter-limit
effect is so stark and clean, it makes sense to correct for it using
$V/V_{\rm max}$ weighting. I compute the weights $w$ for each galaxy as
\begin{eqnarray}
w & = & V_{\rm tot} / V_{\rm max} \;\; \mathrm{if} \; V_{\rm max} < V_{\rm tot}, \\
  & = & 1 \;\; \mathrm{otherwise,}
\end{eqnarray}
where $V_{\rm tot}$ is the total volume out to the distance limit
$D_{\rm lim}$ (e.g., $D_{\rm lim} =25$ Mpc for Sample~1) and $V_{\rm
max}$ is the volume enclosed by $D_{\rm max}$, the maximum distance
at which the galaxy could still have exceeded the survey's diameter
limit. $D_{\rm max}$ depends on the galaxy's $R_{25}$ in kpc and thus
on the observed $R_{25}$ angular size and the galaxy's actual distance $d$;
the weights can thus be expressed as
\begin{eqnarray}
w & = & (D_{\rm max}^{3} \, R_{25, \mathrm{lim}}) / (d^{3} \, R_{25}) \;\;  \mathrm{if} \; D_{\rm max} < D_{\rm lim}, \\
  & = & 1 \;\; \mathrm{otherwise,}
\end{eqnarray}
where $R_{25, \mathrm{lim}} = 30\arcsec$.

\begin{table}
\caption{\sfourg-based Samples}
\label{tab:samples}
\begin{tabular}{lcrr}
\hline
Name      & $D_{\rm max}$ & Minimum \Mstar{} & $N$ \\
          & (Mpc)         & ($\logmstar$)   &     \\
\hline
Parent Disc Sample  & ---    & ---             & 1322 \\
Parent Spiral Sample  & ---  & ---             & 1220 \\
Sample 1     &  25           & ---             &  659 \\
Sample 1m    &  25           & 8.5             &  574  \\
Sample 2     &  30           & ---             &  851  \\
Sample 2m    &  30           & 9               &  638  \\
\end{tabular}

\medskip

Definitions and galaxy counts for different subsamples of \sfourg{} galaxies
used in this paper. Columns: (1) Subsample name. (2) Distance upper limit. (3)
Lower limit on stellar mass. (4) Number of galaxies in subsample.
\end{table}

\section{Bar Fractions for Local Galaxies}\label{sec:fbar} 

The bar fractions for \sfourg{} galaxies from Samples~1 or 1m are shown in
Figure~\ref{fig:fbar-vs-all}, first for for all bars (red circles, left
panels) and then for for strong and weak bars considered separately
(black and cyan circles, right panels). I show these fractions as a
function of three galaxy parameters: stellar mass (top panels), \gmr{}
colour (middle panels), and gas mass ratio \fgas{} (bottom panels). The
error bars attached to each bar fraction account for the $V/V_{\rm max}$ weights 
(and $B_{\rm tc}$-based weights for the \gmr{} plots) using the
approach of \citet{wilman-erwin12}: all weighted counts in a given
bin were rescaled so that the total was equal to the original unweighted
total counts for that bin, and these were then used to estimate the
\citet{wilson27} 68\% confidence limits.  Additional bar-fraction estimates from a
subset of different SDSS-based studies are shown in the left-hand
panels; see below for the specific studies.

\subsection{Bar Fraction and Galaxy Stellar Mass}\label{sec:fbar-mass-s4g} 

\begin{figure*}
\begin{center}
\hspace*{-5mm}\includegraphics[scale=0.85]{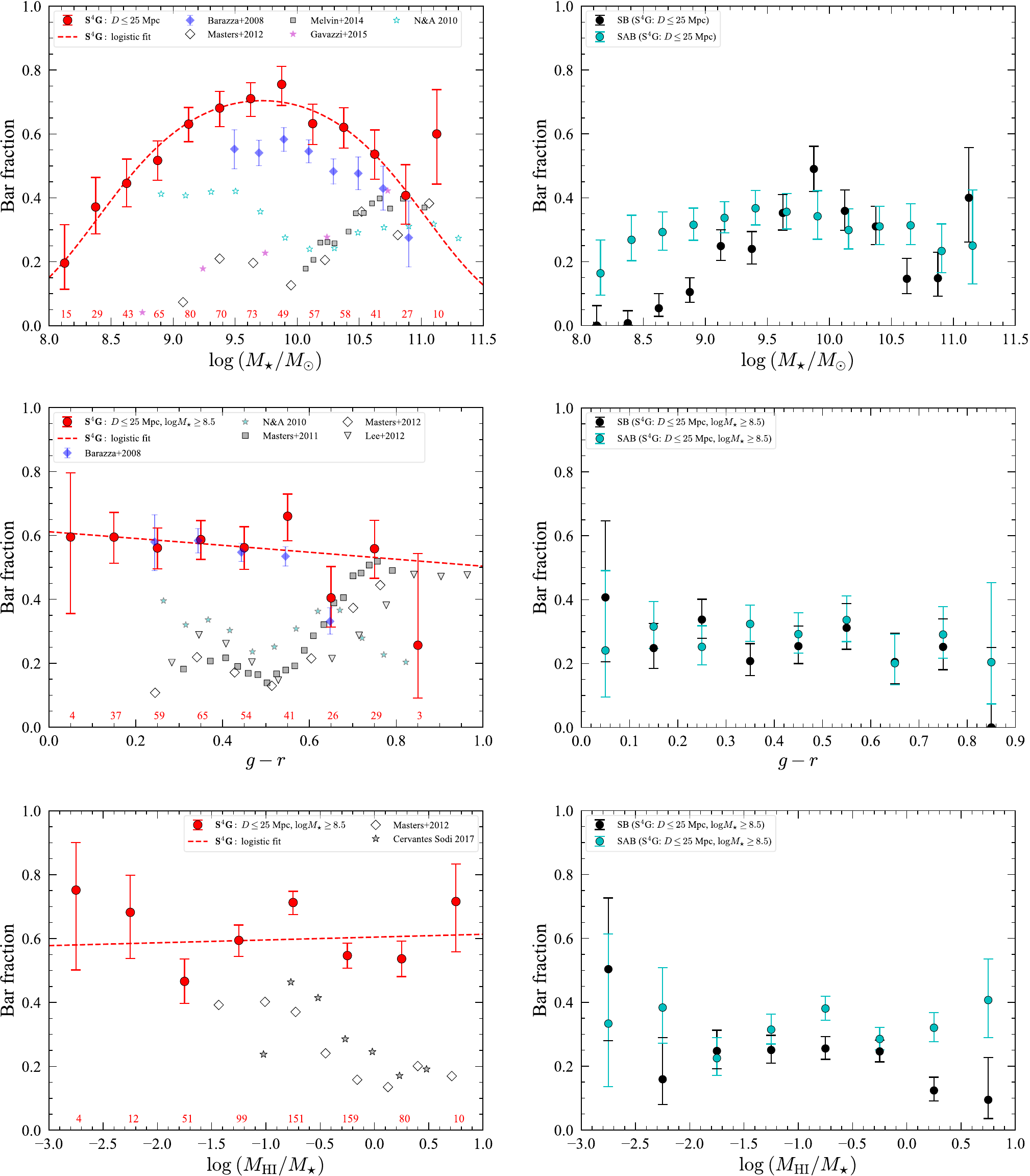}
\end{center}

\caption{Fractions of disc (spiral + Irr) galaxies with bars (left
panels) or with strong (SB) and weak (SAB) bars counted separately
(right panels). Fractions for \sfourg{} samples are shown with red
circles (left panels) or black and green circles (right panels); dashed
red lines show the results of logistic fits to the underlying data; see
text for details. The small red numbers along the bottom of each
left-hand panel are the raw \sfourg{} counts per bin. Top panels: Bar
fractions as a function of stellar mass for \sfourg{} galaxies with
distances $\le 25$ Mpc. Other symbols show reported bar fractions for
GZ2 \citep{masters12,melvin14} and three other SDSS-based studies
\citep{barazza08,nair10b,gavazzi15}. Middle panels: Bar fractions as a
function of \gmr{} colour for \sfourg{} galaxies in Sample 1m (based on
\bmv{} colours). Gray squares and diamonds show reported bar fractions
for GZ2 \citep{masters11,masters12}, with grey triangles for
\citet{lee12a} and cyan stars for \citet{nair10b}. Bottom panels: Bar
fractions as a function of atomic gas fraction \fgas{} for \sfourg{}
galaxies in Sample 1m. Also plotted are reported SDSS-based trends for
GZ2 \citep[][open diamonds]{masters12} and \citet[][gray
stars]{cervantes-sodi17}.\label{fig:fbar-vs-all}}

\end{figure*}

The upper left panel of Figure~\ref{fig:fbar-vs-all} shows how the bar
fraction \fbar{} in \sfourg{} galaxies varies with stellar mass. There
is a steep increase from very low stellar masses, reaching a maximum of
$\fbar = 0.76$ at $\logmstar \sim 9.8$, and then a similarly steep
decline to higher masses. Although Sample~1 is incomplete for $\logmstar
\la 8.5$, this is a volume effect and should not affect the
detectability of bars in lower-mass galaxies; it only reduces the total
number of galaxies and makes the bar fraction more uncertain. Thus, the
steep decline in \fbar{} seen for $\logmstar \la 8.5$ is probably real.

This pattern is very similar to the ``visual'' \fbar{} trend in
\citet[][their Fig.~19; see also
Appendix~\ref{sec:app-subsamples}]{dg16a}, since both are based on the
\citet{buta15} bar classifications for \sfourg. The main difference is
their shallower slope for higher masses ($\logmstar \ga 10$) and the
existence of a secondary peak in \fbar{} at $\logmstar \sim 10.5$. These
differences are (necessarily) due to the fact that Sample~1 is a
distance-limited subset of the Parent Disc Sample, and the fact that it
excludes S0 galaxies. For example, adding the S0s to Sample~1 produces a
flatter trend of \fbar{} versus \logmstar{} for higher stellar masses
(see Appendix~\ref{app:S0s}).

In an attempt to parameterize the trend -- and to avoid dependencies on
the binning used for the figure -- I also plot the results of a polynomial
logistic regression applied to Sample~1 (dashed red line). Logistic regression involves
modeling the presence or absence of a feature (in this case, bars) using a function
which represents the binomial probability
of that feature. It has the advantage of using \textit{all} the data directly
rather than relying on the details of a binning scheme. A generalized polynomial 
version (of order $n$) is
\begin{equation}\label{eq:logistic1}
\fbar(x) \; = \; \frac{1}{1 + e^{-(\alpha + \sum_{i = 1}^{n} \beta_{i} x^{i})}} ,
\end{equation}
where $x$ is the independent variable (e.g., \logmstar). Although
logistic regression typically involves a linear function of $x$, leading
to a probability that either monotonically increases or decreases in a
sigmoid fashion, the behavior of \fbar{} is clearly \textit{not}
monotonic, so a quadratic function ($n = 2$) is probably better. The
result (with best-fit coefficients $\alpha = -82.2 \pm 22.1$,
$\beta_{1} = 17.1 \pm 4.62$, and $\beta_{2} = -0.88 \pm 0.24$) is
plotted as a dashed red line in Figure~\ref{fig:fbar-vs-all}; the peak
in \fbar{} is $\approx 0.70$ at $\logmstar \approx 9.7$. (The fitting
was done to data from Sample~1 from $\logmstar =
8$--12 with $V/V_{\rm max}$-based weights, using the R Survey
package.\footnote{\url{http://r-survey.r-forge.r-project.org/survey/index.html}})

Also shown in the upper left panel of Figure~\ref{fig:fbar-vs-all} are
the reported trends for GZ2 from \citet{masters12} and (for a larger
sample) from \citet{melvin14}. Both of these studies suggest a very low
\fbar{} for $\logmstar \la 10$ and a steep increase in \fbar{} for
higher stellar masses, with a maximum at $\logmstar \sim 10.8$.
A similar trend can be seen in the non-GZ2 SDSS-based results of
\citealt{gavazzi15}. This clearly does \textit{not} agree with the
\sfourg{} values; indeed, for $\logmstar \sim 10$--11, the \sfourg{} and
SDSS-based bar frequencies show the \textit{opposite} behavior. The more
complicated dependence seen by \citet{nair10b} is not apparent in the
\sfourg{} data, either.

There \textit{is} rather good agreement with the results of
\citet[][blue diamonds]{barazza08},\footnote{For the \citet{barazza08} results
here and in the left middle panel, I use their per-bin total-galaxy counts
to estimate proper binomial uncertainties; this is generally not possible for
the other SDSS-based studies.} and fairly good agreement with
\citet[][not shown]{mendez-abreu12}, especially with the latter's
``Field'' subsample, where \fbar{} peaks at $\logmstar \sim 9.5$;
\fbar{} for their ``Virgo'' subsample peaks at $\logmstar \sim
10.1$. Since the Virgo subsample is at distances $< 20$ Mpc, and since most
of the field subsample was defined with redshifts between 2500 and 3000 \kms,
the spatial resolution of the SDSS images they use is generally similar
to that of the \sfourg{} sample.

The upper right panel of Figure~\ref{fig:fbar-vs-all} shows the bar
fraction separately for strong (SB) and weak (SAB) bars. The SAB
fraction is almost constant for $\logmstar \sim 8.5$--10.7,
declining slightly to both lower and higher masses. Strong bars show a more
dramatic version of the global bar trend: \fbar{} increases from
$\sim 0$ at the lowest stellar masses to a rather sharp peak at
$\logmstar \approx 9.8$, and then falls off steeply to higher masses.
This suggests that the clear dependence of \fbar{} on stellar mass is
driven primarily by \textit{strong} bars.

\subsection{Bar Fraction and Galaxy Colour}\label{sec:fbar-color-s4g} 

The middle panels of Figure~\ref{fig:fbar-vs-all} shows the fraction of
\sfourg{} galaxies in Sample~1m with bars, as a function of \gmr{}
colour,\footnote{Corrected for colour incompleteness as described in
Section~\ref{sec:color-incompleteness}.} for all bar types (left panel)
and separately for strong and weak bars (right panel). The overall relation
is fairly flat, although it is possible to make out a weak trend of
\fbar{} declining to redder colours. The dashed red line shows a standard
(linear) logistic regression to the underlying data; it indicates a declining
bar fraction to redder colours, though the nonzero slope is not statistically
significant. Figure~\ref{fig:fbar-vs-BmV} (in
Appendix~\ref{sec:fbar-vs-BmV}) shows almost identical behavior when
\bmv{} colour is used instead of \gmr.

I also plot the published GZ2 trends from \citet{masters11} and
\citet{masters12} and the non-GZ2 trend from \citet{lee12a} in the middle
left panel (grey symbols). These disagree strongly with the \sfourg{}
trend, particularly given the steep rise in GZ2 \fbar{} to redder colours
($\gmr \sim 0.5$--0.75).

The trend of \fbar{} versus colour from \citet[][cyan stars]{nair10b} is
similar to the GZ2 trend for intermediate colours ($\gmr \sim 0.3$--0.65),
but diverges for the reddest colours. It, too, agrees poorly with the
\sfourg{} trend.

In contrast to the other SDSS-based studies, the \citet{barazza08} trend
(blue diamonds) shows extremely good agreement with the \sfourg{} results,
including the slight decrease in \fbar{} to redder colours. 

The separate trends for strong and weak bars (right-hand middle panel) are, within
the uncertainties, indistinguishable and roughly constant.

\subsection{Bar Fraction and Atomic Gas Fraction}\label{sec:fbar-fgas-s4g} 

In the lower left panel of Figure~\ref{fig:fbar-vs-all} I plot \fbar{}
as a function of the atomic gas mass ratio \fgas. The \sfourg{} bar fraction is
basically constant, apart from a weak minimum at $\logfgas \sim -1.3$.
The dashed red line shows a standard (linear) logistic regression to the
underlying data; the slope is not statistically significant, so there is
no evidence for a change in \fbar{} with \logfgas.

Plotted in the same panel is the GZ2 trend reported by
\citet{masters12}, which shows fairly strong disagreement: the GZ2
\fbar{} value drops steeply for \logfgas{} between $-1.0$ and 0, while
for \sfourg{} galaxies \fbar{} reaches a (weak) local maximum within the
same range, and shows no sign of a systematic decline. Also shown are
the reported \logfgas{} values from the SDSS-based study of
\citet{cervantes-sodi17},\footnote{These have had 0.146 dex subtracted
to account for the fact that Cervantes Sodi scaled the \hi{} gas masses
by a factor of 1.4 in order to include helium.} which show a pattern
very similar to the GZ2 one.

The lower right panel of Figure~\ref{fig:fbar-vs-all} shows how the
frequencies of strong and weak bars behave as function of gas mass
ratio. For values of \logfgas{} between $-2$ and 0.5 (\fgas{} between
$\sim 0.01$ and 0.3), the fractions of both strong and weak bars are
essentially identical -- and constant. This may also be true for very low
gas fractions, though the small numbers make the fractions uncertain.
Only for very high gas fractions ($\fgas \ga 1$) is there a
noticeable difference: the fraction of weak bars increases, while the
fraction of strong bars goes down.

It is important to note here that the SDSS-based studies are
\textit{not} directly comparable in terms of sample definition.
Specifically, the sample of \citet{masters12} was defined using \hi{}
detections from the $\alpha 40$ release of Arecibo Legacy Fast ALFA
(ALFALFA) survey \citep{giovanelli05,haynes11}. Only 51\% of the GZ2
galaxies observed as part of ALFALFA were detected in \hi; the resulting
gas mass fractions are relatively high (median $\fgas = 0.39$ for their
barred galaxies and 0.74 for the more numerous unbarred galaxies). This
contrasts strongly with the $\sim 97$\% \hi{} detection rate for the
\sfourg{} sample; the median value of \fgas{} for the \sfourg{} galaxies
is 0.25. (If we match the GZ2 sample more closely by only considering
\sfourg{} galaxies with $\logmstar \ge 9$, then the \sfourg{} median is
even lower: $\fgas = 0.19$.) Similar considerations undoubtedly apply to
the \fbar{} versus \fgas{} analysis of \citet{cervantes-sodi17}, since
that study used an expanded version of the same ALFALFA \hi{}
observations and a similar set of SDSS galaxies.

Since the \sfourg{} sample is far more complete in terms of \hi{} detections,
and samples down to much lower values of \fgas, than is true for the
SDSS studies, it is probably more representative in terms of how bars do
(or do not) depend on the gas mass fraction in galaxies.

\begin{table*}
 \centering
 \begin{minipage}{140mm}
\caption{Total Bar Fractions}
\label{tab:fractions}
\begin{tabular}{lcrlll}
\hline
Name      & $D_{\rm max}$ & Minimum \Mstar{} & \fbar{} & \fSB{} & \fSAB{} \\
          & (Mpc)         & ($\logmstar$)    &     \\
\hline
Sample 1     &  25           & ---           & $0.563 \pm 0.019$ & $0.234^{+0.017}_{-0.016}$ & $0.329 \pm 0.018$ \\
Sample 1m    &  25           & 8.5           & $0.618 \pm 0.020$ & $0.267^{+0.019}_{-0.018}$ & $0.351 \pm 0.020$ \\
Sample 2     &  30           & ---           & $0.564 \pm 0.017$ & $0.236^{+0.015}_{-0.014}$ & $0.328 \pm 0.016$ \\
Sample 2m    &  30           & 9             & $0.623 \pm 0.019$ & $0.290 \pm 0.018$         & $0.333^{+0.019}_{-0.018}$ \\
\hline
\end{tabular}

\medskip

Observed bar fractions for different subsamples of \sfourg{} spiral galaxies.
For each subsample, I list the total bar fraction \fbar{} and the separate
fractions for strong (SB) and weak (SAB) bars. The fractions for Samples~1 and 2
include all galaxies within the specified distance limits; the fractions for
Samples~1m and 2m are restricted to galaxies more massive than the specified
stellar-mass limits.

\end{minipage}
\end{table*}

\subsection{Strong Bars versus Weak Bars}\label{sec:strong-bars} 

Several SDSS-based papers have argued that their detections are
primarily of ``strong'' bars. For example, by comparing galaxies in
common between GZ2 and the sample of \citet{nair10b}, \citet{masters12}
suggested that galaxies with GZ2 $p_{\rm bar} \ge 0.5$ -- their standard
definition for a galaxy being barred -- were equivalent to the ``strong
bars'' of \citet{nair10a,nair10b}. Is there any evidence that local
(\sfourg) galaxies behave more like GZ2 and other SDSS galaxies when
only strong bars are considered?

Figure~\ref{fig:fbar-vs-all} suggests that strong (SB) and weak (SAB)
bars in \sfourg{} behave almost indistinguishably, with two exceptions.
First, strong bars display a more sharply peaked distribution as a
function of stellar mass, with a maximum at $\logmstar \sim 9.8$, in
contrast to the very broad maximum in \fbar{} seen for SAB bars (upper
right panel of Figure~\ref{fig:fbar-vs-all}). Second, the fractions of
SB and SAB bars (\fSB{} and \fSAB) diverge for very high gas mass
ratios: for $\fgas > 1$, \fSB{} decreases and \fSAB{} increases. But
neither of these SB trends agrees with the general SDSS-based results.
Indeed, the \fSB{} trend as a function of stellar mass is if anything
even \textit{more} different from the SDSS-based trends than is true for
\fbar.

Similarly, the fact that the \sfourg{} \fSB{} and \fSAB{} values remain
roughly constant (and equal) for $\logfgas < 0$ (lower right panel of
Figure~\ref{fig:fbar-vs-all}) means that the strong change in \fbar{}
versus \logfgas{} seen by \citet{masters12} and \citet{cervantes-sodi17}
(over the range $\fgas \approx -1.0$ to 0) cannot be due to their
detecting only strong bars. Finally, the trends of \fSB{} and \fSAB{} as
a function of \gmr{} (middle right panel of
Figure~\ref{fig:fbar-vs-all}) are identical within the uncertainties,
with both indicating either no correlation with colour or else a slight
decrease in bar fraction towards redder colours.

So it appears that we cannot explain the discrepancies between the
\sfourg{}-based results and the SDSS-based results by appealing to the
hypothesis that the latter are simply the behavior of \textit{strong}
bars rather than bars in general. 

It should be kept in mind that the concept of ``strong'' versus
``weak'' bars is inherently a somewhat ambiguous one. Various
quantitative measures of bar strength have been proposed -- and several
of them have been applied to many (though not all) of the bars in
\sfourg{} -- but they can sometimes be surprisingly contradictory, in
part because they aim at defining different aspects of ``strength''.
Figs.~17, 18, and 20 of \citet{dg16a} shows some of the potential
ambiguity for the \sfourg{} galaxies. For example, if one chooses the
\atwomax{} measure often used by theorists, then there is fairly clear
correlation (albeit with considerable scatter) between galaxy mass and
bar strength: on average, the strongest bars are in high-mass galaxies.
If one uses the maximum isophotal ellipticity (probably the easiest
observational measurement), then there is \textit{no} correlation. For
$Q_{b}$ (the maximum gravitational torque produced by the bar), there is
an \textit{anti}-correlation, with the strongest bars in lower-mass
galaxies.\footnote{This is at least partly because $Q_{b}$ measures the
gravitational strength of the bar relative to the bulge and inner disk;
in low-mass galaxies with little or no bulge, the bar's impact is less
diluted and so $Q_{b}$ is stronger.}  Similarly, if one follows
\citet{lee12a} and \citet{cervantes-sodi17} and uses the size of the bar
relative to the apparent or fiducial size of the disc, then the
strongest bars would be in the \textit{lowest}-mass galaxies (see also
Figure~\ref{fig:barsize-R25}). In general, one should probably be
cautious about claims that SDSS-based studies are detecting mostly (or only) ``strong
bars'', since unless the \textit{type} of bar strength is carefully
specified, the implications are unclear.

\subsection{The Global Bar Fraction} 

As demonstrated in the preceding subsections, the bar fraction \fbar{}
is roughly constant with both colour and gas mass ratio, but is a strong
function of stellar mass. The latter fact means that there is no one
``global'' bar fraction, since the fraction for any sample will depend
on the masses of the galaxies in the sample. With that caveat in mind, I
present summary fractions in Table~\ref{tab:fractions}, which shows the
bar fraction for the different \sfourg{} spiral subsamples. The bar
fraction for \textit{all} spirals is $0.56 \pm 0.02$; however, this is
probably a slight \textit{overestimate} given the incompleteness in
galaxies with $\logmstar < 9$, since low-mass galaxies are less likely
to have bars. For spirals with $\logmstar \ga 9$, the fraction is $0.62
\pm 0.02$. This is consistent with most previous estimates from smaller
local samples (e.g., \citealt{sellwood93,
mulchaey97,knapen00,eskridge00,menendez-delmestre07}; and the discussion
in \citealt{sheth08}), which tended to be dominated by high-mass
galaxies.

\section{The Sizes and Detectability of Bars}\label{sec:bar-sizes} 

\subsection{Why Do \sfourg{} Bar Frequencies Differ From (Most) SDSS Results?} 

The previous section has shown that the frequencies of bars in
local (\sfourg) galaxies generally differ markedly from those reported
for most SDSS-based samples -- not just in overall frequency, but also
in trends with stellar mass, colour, and gas fraction. Why is this so?

It seems unlikely that cosmic variance could produce such a strong,
systematic difference. The fact that the \citet{barazza08} and
\citet{mendez-abreu12} studies actually agree rather rather well with
the local results makes such an appeal even more improbable. Similarly,
while the use of near-IR images for the \sfourg{} sample largely
eliminates the possibility of missing bars due to strong dust extinction
or star formation, which might affect SDSS optical images, it is unclear
why dust and star formation should \textit{not} affect the
\citet{barazza08} or \citet{mendez-abreu12} results in the same way as
most other SDSS-based studies. Differences in bar-detection
\textit{methods} could certainly explain \textit{some} of the
difference, particularly since some methods may be more sensitive to
weaker or smaller bars than others. Still, SDSS-based studies with
different bar-detection methods \textit{do} show similar results: e.g.,
GZ2 \citep{masters11,masters12,skibba12}, \citet{oh12}, and
\citet{lee12a} for \fbar{} versus \gmr; GZ2
\citep{masters12,skibba12,melvin14}, \citet{oh12}, \citet{gavazzi15},
and \citet{cervantes-sodi17} for \fbar{} versus stellar mass; or GZ2 and
\citet{cervantes-sodi17} for \fbar{} versus gas mass fraction. So it is
hard to see how the strong differences between most SDSS-based studies
and the local, \sfourg{} results can be explained this way.

One key difference between the \sfourg{} galaxies and the SDSS samples
is \textit{spatial resolution}. Although the FWHM for the
\textit{Spitzer} 3.6\micron{} images used for \sfourg{} ($\approx 2\arcsec$) is
slightly worse than the typical SDSS seeing of $\sim 1.4\arcsec$, SDSS
galaxies are, on average, much further away, so the effective linear
resolution is correspondingly much poorer. The mean distance of \sfourg{}
galaxies in Sample~1 is 16.8 Mpc (median = 17.8 Mpc), so the FWHM
translates to a mean linear resolution of $\sim 165$~pc. For a
volume-limited sample spanning $z = 0.01$--0.06, as used by most of the GZ2
studies, the mean redshift would be $\approx 0.045$ (assuming a uniform
galaxy density), with a mean linear resolution of $\approx
1.25$~kpc -- almost an order of magnitude worse.

In this context, it is significant that the SDSS-based studies which
\textit{do} show agreement with the local galaxies -- that is,
\citet{barazza08} and \citet{mendez-abreu12} -- are samples of galaxies
at smaller distances. The \citet{barazza08} sample had a redshift range
of $z = 0.01$--0.03, which implies (assuming uniform densities) $\langle
z \rangle \approx 0.023$ and a mean spatial resolution twice as good as
GZ2 and similar SDSS-based studies. The ``field2'' subsample of
\citet{mendez-abreu12} used galaxies with redshifts of 2500--3000
\kms{}, their ``Virgo'' subsample used Virgo Cluster galaxies, and their
``Coma'' subsample used \textit{HST} rather than SDSS images, so their
spatial resolutions were similar to or slightly better than that of the
\sfourg{} galaxies.

Poorer resolution has the obvious effect of reducing the observed
ellipticity and contrast of a bar (perhaps especially so if there is a
bright bulge whose light can be convolved with that of the bar), and
this will be worse for bars which are smaller in angular size. 
Figure~\ref{fig:barsize-vs-distance} suggests a possible limit of $\sim
2 \times {\rm FWHM}$ for detecting bars in \sfourg. Although there is a
potential complication due to the existence of an angular size limit for
the parent \sfourg{} sample -- e.g., at larger distances galaxies with
smaller optical discs are preferentially excluded, which may
in turn exclude smaller bars -- \citet{aguerri09} concluded from
their experiments with artificial galaxy images that a lower limit on
detectable bar size for SDSS images was $a_{{\rm bar}} \sim 3.6\arcsec$,
or about $2.5 \times {\rm FWHM}$.

In the context of GZ2, \citet{willett13} explicitly pointed out a
possible bias against detecting smaller bars, based both on the GZ2
public interface (``GZ2 participants may not have looked for bars
shorter than the disc diameter, or have been less confident in voting
for `yes' if they were identified'') and also on a strong correlation
between the GZ2 $p_{\rm bar}$ parameter and the EFIGI \citep{baillard11}
bar ``length'' parameter for galaxies in common between GZ2 and EFIGI.

So -- despite arguments sometimes made that all ``normal'' or
``galactic-scale'' bars can be detected in SDSS samples -- it is at
least possible that most SDSS studies are failing to detect significant
numbers of bars due to resolution effects. (Some SDSS-based studies --
e.g., \citealt{gadotti11,wang12} -- \textit{are} careful to point out
that they are less sensitive to smaller bars.) Whether this might
explain the different systematics in bar frequency, depends on, among
other things, the actual sizes of bars -- and whether their sizes vary
significantly with galaxy properties such as mass, colour, or gas
content.

\subsection{The Sizes of Bars in \sfourg{}} 

\begin{figure}
\begin{center}
\hspace*{-2.5mm}\includegraphics[scale=0.46]{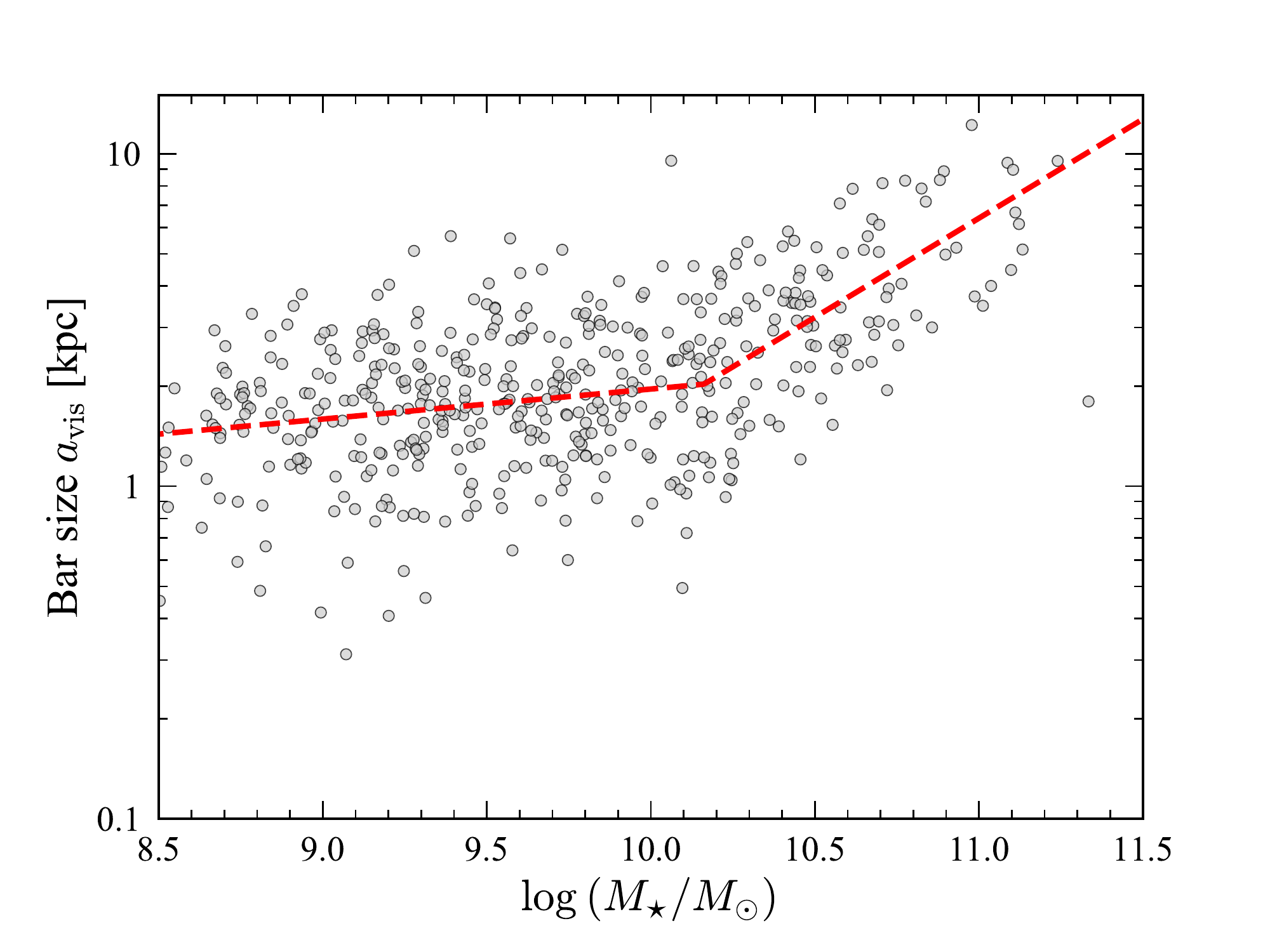}
\end{center}

\caption{Bar semi-major axis versus stellar mass for \sfourg{} spiral
galaxies with $D \leq 25$ Mpc (dark circles); galaxies with distances
between 25 and 30 Mpc are plotted with open circles. The red dashed line
shows a broken linear fit to galaxies with $D \leq 30$ Mpc and $\logmstar =
9$--11; see Paper~II for more details.\label{fig:barsize-with-fit}}

\end{figure}

Figure~\ref{fig:barsize-with-fit} shows how bar size behaves as a
function of stellar mass for \sfourg{} galaxies in Samples~1 and 2. There
are two important things to note about this figure. 

The first is that there is considerable scatter in intrinsic bar
size for any given stellar mass; among other things, this scatter means
that a significant fraction of bars have semi-major axes $< 1$ kpc.
Although many studies of bars in SDSS samples correctly note that bars
with sizes $\la 1$ kpc will probably go undetected in SDSS images, they
often argue that bars of this size are ``nuclear bars'' distinct from
``normal'' bars, and thus any failures to reliably detect bars of this
size do not affect conclusions about bars (or normal bars) in general.
But as Figure~\ref{fig:barsize-with-fit} demonstrates, sub-kpc bars
appear to be simply the tail end of the general distribution of bar
sizes for galaxies with masses $\logmstar \la 10.2$, and so failing to
detect such bars \textit{will} bias conclusions about bar frequencies.
Only for galaxies with masses $\logmstar \ga 10.3$ could one argue that
bars with lengths $< 1$ kpc would be different from the general
population of bars.

The second point is that there is indeed, as noted by \citet[][see their
Fig.~20 and Table~5]{dg16a}, a clear correlation between bar size and
stellar mass. Closer inspection suggests the trend actually has a
\textit{bimodal} quality: bar size is almost constant (but increases
slightly with stellar mass) for stellar masses $\la 10^{10} \Msun$, and
clearly increases with stellar mass for higher masses.

Further analysis of the correlation between bar size and stellar mass
(and other galaxy properties) will be presented in Erwin (2018, in
prep; hereafter Paper~II). Here, I emphasize the basic size-mass
correlation, as well as the general observation that a correlation
between bar size and stellar mass will probably translate into a
correlation between bar size and galaxy colour and an anticorrelation
between bar size and gas content (since more massive galaxies tend to
redder and more gas-poor; see Figure~\ref{fig:gmr-and-fgas-vs-mstar}).
Evidence for this (especially for the gas-content--bar-size
anticorrelation) can be seen in
Figure~\ref{fig:barsize-vs-all-with-fwhm}.

\begin{table*}
\begin{minipage}{126mm}
\caption{Fits to Bar Size versus Stellar Mass}
\label{tab:bar-size-fits}
\renewcommand{\arraystretch}{1.5}
\begin{tabular}{rcccccc}
\hline
Fit range    & $D_{\rm max}$ & $\alpha_{1}$ & $\beta_{1}$ & $\alpha_{2}$  & $\beta_{2}$ & $\log \: (M_{\rm brk} / \Msun)$ \\
(\logmstar)  &  (Mpc)        &              &             &               &             &        \\
  (1)        & (2)           & (3)          & (4)         & (5)           &        (6)  & (7)    \\
\hline
$\geq 8.5$ & 25         & $-0.62^{+0.36}_{-0.22}$   & $0.09^{+0.02}_{-0.04}$    & $-4.35^{+1.08}_{-0.55}$   & $0.46^{+0.05}_{-0.10}$    & $10.01^{+0.04}_{-0.16}$   \\
8.5--11 & 25         & $-0.75^{+0.26}_{-0.30}$   & $0.10^{+0.03}_{-0.03}$    & $-5.80^{+0.87}_{-0.92}$   & $0.60^{+0.09}_{-0.08}$    & $10.16^{+0.08}_{-0.07}$   \\
$\geq 9$ & 30         & $-0.69^{+0.28}_{-0.52}$   & $0.10^{+0.05}_{-0.03}$    & $-4.56^{+0.72}_{-0.56}$   & $0.48^{+0.05}_{-0.07}$    & $10.10^{+0.09}_{-0.06}$   \\
9--11 & 30         & $-0.53^{+0.41}_{-0.38}$   & $0.08^{+0.04}_{-0.04}$    & $-5.39^{+1.13}_{-0.47}$   & $0.56^{+0.05}_{-0.11}$    & $10.11^{+0.02}_{-0.13}$   \\
All & ---        & $-1.17^{+0.22}_{-0.31}$   & $0.15^{+0.03}_{-0.02}$    & $-4.17^{+0.61}_{-0.66}$   & $0.44^{+0.06}_{-0.06}$    & $10.24^{+0.07}_{-0.03}$   \\
\hline
\end{tabular}

\medskip

Results of broken-linear fits to $\log \avis$ as a function of \logmstar.
Parameter uncertainties are based on 2000 rounds of bootstrap resampling.

\end{minipage}
\end{table*}

\subsection{Implications for SDSS-Based Surveys} 

The implication of the bar-size--stellar-mass correlation is this:
because higher-mass galaxies tend to have physically larger bars, their
bars will be easier to detect at large distances than is true for bars
in lower-mass galaxies. If the resolution limit of a survey is well
below the size of most or all of the bars in a sample, even for the most
distant galaxies -- as is the case for the distance-limited subsamples
of \sfourg{} -- then very few bars will be missed. But most SDSS-based
studies have a typical linear resolution for their galaxies that is
about an order of magnitude worse than that of \sfourg{}. They will
accordingly be much more vulnerable to resolution effects. And since the
resolvability of bars depends on both survey resolution and their linear
sizes, bars in lower-mass galaxies will be more easily missed than those in
high-mass galaxies.
 
\begin{figure}
\begin{center}
\hspace*{-2.0mm}\includegraphics[scale=0.67]{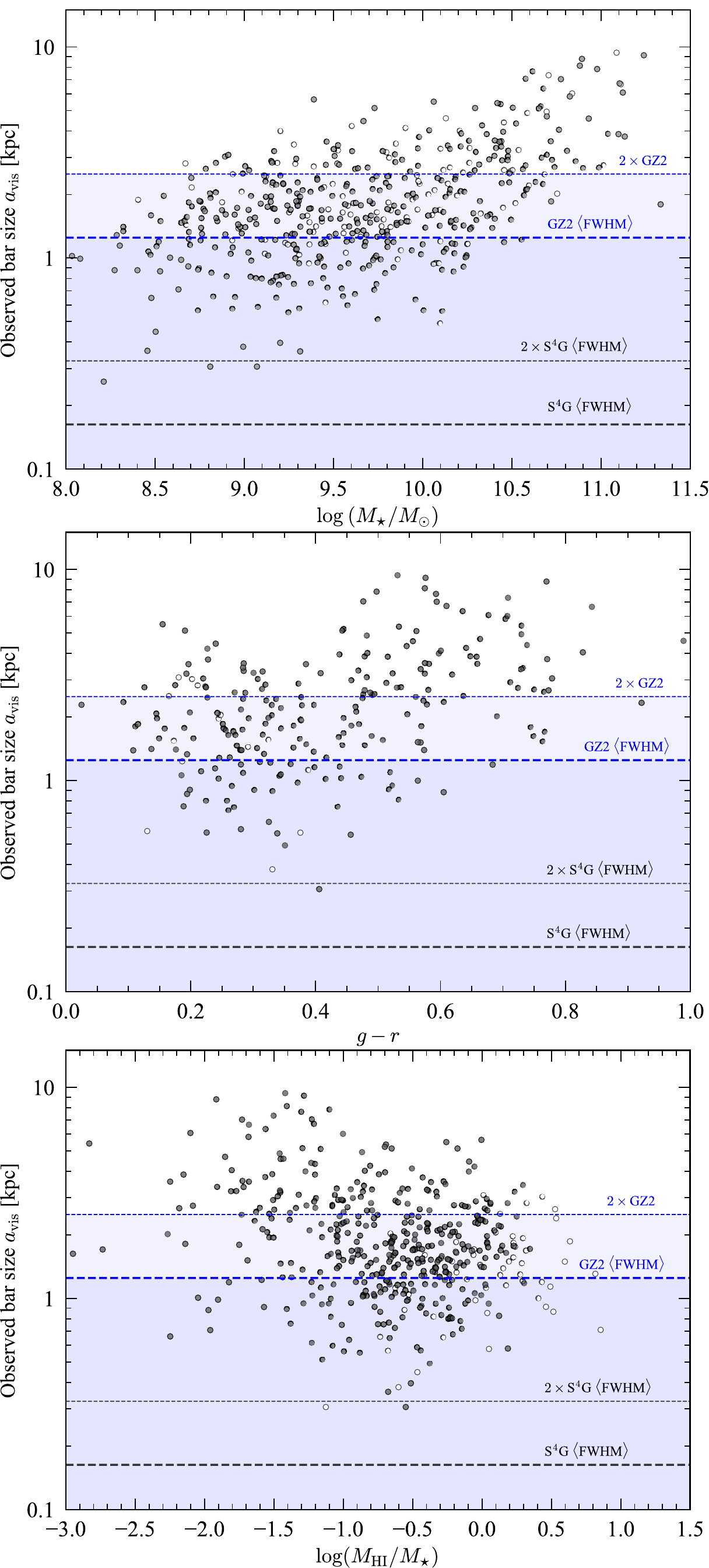}
\end{center}

\caption{Observed (not deprojected) sizes of bars in \sfourg{}
galaxies versus stellar mass (top), $g - r$ colour (middle), and gas mass
ratio (bottom). Filled circles indicate galaxies in Sample~1 ($D \leq
25$ Mpc, top panel) or Sample~1m ($D \leq 25$ Mpc, $\logmstar \geq 8.5$,
middle and bottom panels); open circles show additional galaxies with $D
= 25$--30 Mpc and also $\logmstar \geq 9$ (middle and bottom panels).
Horizontal dashed lines show typical linear resolution for galaxies in
\sfourg{} (black, FWHM $= 2\arcsec$, $\langle D \rangle = 17$ Mpc for
Sample~1) and in GZ2 or similar SDSS-based studies (blue, FWHM $=
1.4\arcsec$, $\langle z \rangle \approx 0.045$). Even assuming that all
bars with angular sizes down to one FWHM are detected, SDSS-based
surveys will miss a significant fraction of bars in lower-mass, blue,
and gas-rich galaxies (dark shaded regions). More bars will be missed if
the actual detection limit is closer to $2 \times \mathrm{FWHM}$ (dark +
light shaded regions).\label{fig:barsize-vs-all-with-fwhm}}

\end{figure}

Figure~\ref{fig:barsize-vs-all-with-fwhm} shows how the different
resolutions interact with the distribution of bar sizes  by plotting
\sfourg{} \textit{observed} bar size \avis{} versus stellar mass, \gmr{}
colour, and \fgas.  (Using the observed sizes helps incorporate the
effects of inclination and projection.) The typical linear resolutions
of both \sfourg{} (165 pc for Sample~1's mean distance of 17 Mpc,
assuming FWHM $\approx 2\arcsec$) and GZ2 (1.25 kpc for $\langle z
\rangle = 0.045$, assuming FWHM $= 1.4\arcsec$) are indicated on the
figure with horizontal black (for \sfourg) and blue (for GZ2) lines.
(``GZ2'' should be understood as representing typical SDSS-based bar
surveys, not just the GZ2 studies.)
Since Figure~\ref{fig:barsize-vs-distance} suggests that a plausible
lower limit for bar detectability is $a \sim 2 \times \textrm{FWHM}$, I
also plot twice the mean FWHM values for each case, using thinner dashed
lines.  The mean \sfourg{} resolution is clearly well below the observed
sizes of almost all the bars. But the mean SDSS resolution excludes a
significant fraction of the bars, and does so in a clearly differential
fashion: more bars are below the resolution limit for lower stellar
masses, for bluer colours, and for higher \fgas{} values. This means
that SDSS-based studies are likely to detect fewer of the bars in
lower-mass, blue, and gas-rich galaxies. (The implied trend is less
clear for bar size versus galaxy colour, with bars apparently becoming
larger in the bluest galaxies. The problem is that these are the
galaxies for which local colour information is most incomplete.)

\subsubsection{Evidence for Missing Small Bars in SDSS Studies}


\begin{figure*}
\begin{center}
\hspace*{-18.5mm}\includegraphics[scale=0.565]{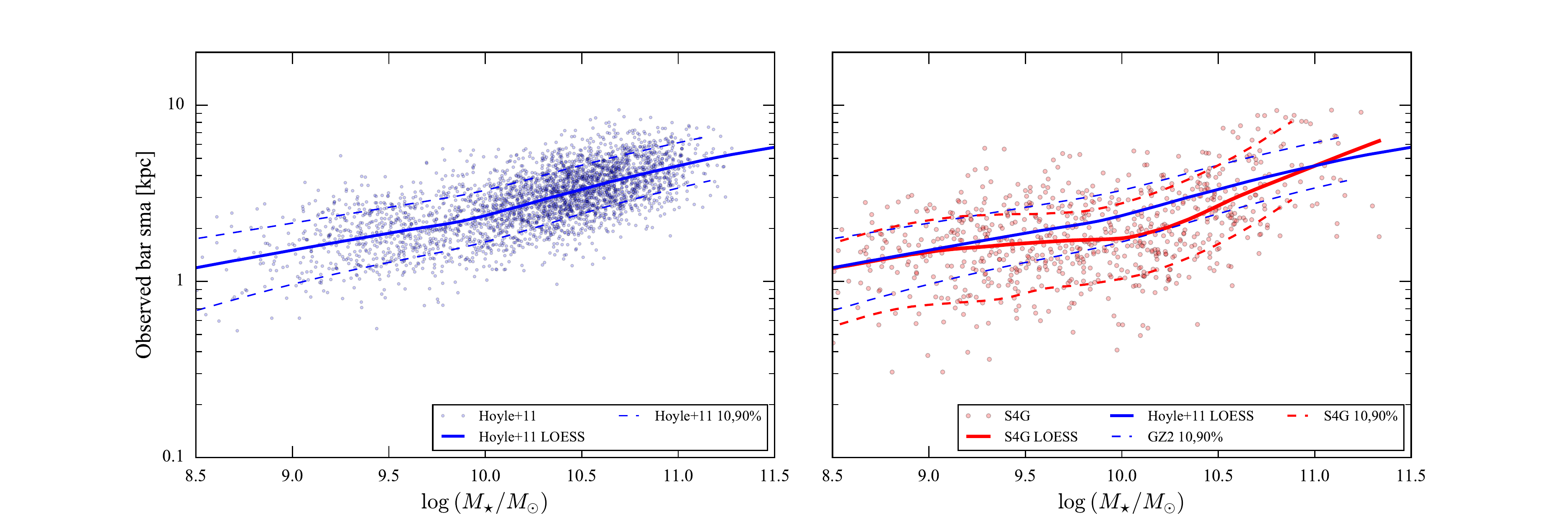}
\end{center}

\caption{Comparison of observed bar sizes between GZ2 and \sfourg. Left:
Observed GZ2 bar sizes from \citet{hoyle11}, as a function of stellar
mass. The solid blue line is a LOESS fit to the data; the dashed lines
outline the 10--90\% quantile range. Right: Same, but now showing
\sfourg{} observed bar sizes, with red lines indicating LOESS and
10--90\% quantiles. The blue lines repeat the GZ2 LOESS and quantile
ranges from the left panel.\label{fig:barsizes-H11-vs-S4G}}

\end{figure*}

Is there more direct evidence that large-scale SDSS-based surveys have
preferentially missed smaller bars? The left panel of
Figure~\ref{fig:barsizes-H11-vs-S4G} compares the observed sizes of bars
as a function of stellar mass from GZ2, using the published bar sizes of
\citet{hoyle11} and stellar masses estimated \citep[using the
colour-to-$M/L$ relations of][]{zibetti09} from $i$-band absolute
magnitudes and $g - i$ colours in the GZ2 catalogs \citep{willett13}.
The solid blue line is a LOESS (locally weighted regression) fit,
somewhat analogous to a running mean \citep[using the implementation
of][]{cappellari13b-atlas3d-20}; the dashed lines outline the 90\%
quantile limits, using the LOESS-based approach of \citet{sakov10}. The
distribution of \sfourg{} bar sizes is shown in the right panel of the
same figure, along with corresponding LOESS fit and 90\% quantiles in red. 

By comparing the overall trends (red lines versus blue lines), we
can see that while the GZ2 and \sfourg{} bars span a similar range of sizes at the
high-mass end (i.e., $\logmstar \sim 11$), the distributions separate as one
goes to lower masses, with the GZ2 distribution biased to larger sizes
relative to the \sfourg{} bars. By the time we reach $\logmstar \sim 10.2$,
almost 90\% of the GZ2 bars are larger than the typical \sfourg{} bar.

\begin{figure}
\begin{center}
\hspace*{-3.5mm}\includegraphics[scale=0.47]{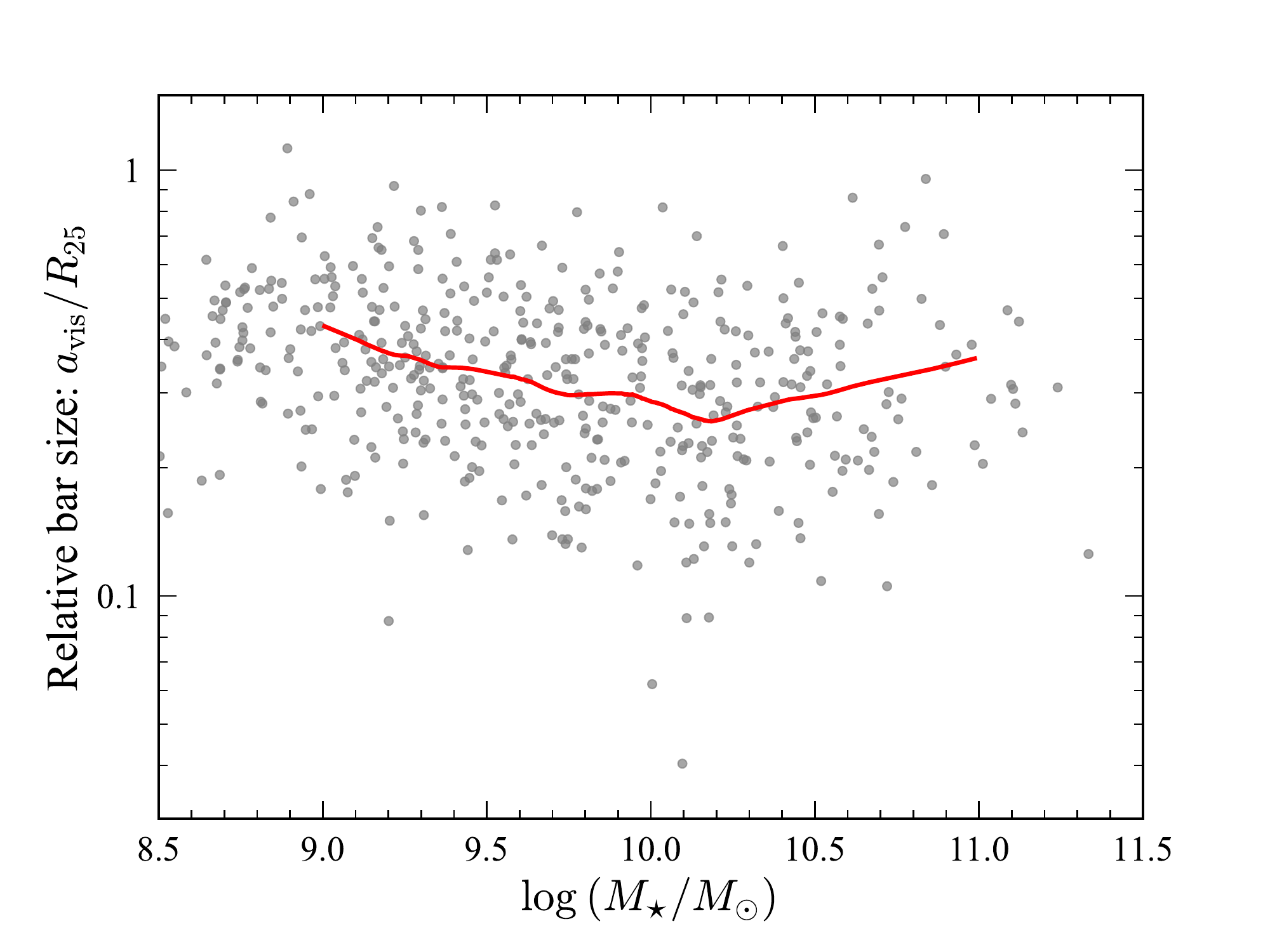}
\end{center}

\caption{Relative bar size (bar semi-major axis divided by blue optical
radius $R_{25}$) for \sfourg{} galaxies in Sample~2. The red line indicates
a LOESS fit to galaxies with $\logmstar = 9$--11. Note that relative
bar size reaches a minimum at $\logmstar \sim 10.2$, and is actually
largest for stellar masses of $\logmstar \sim 9$. See also Fig.~20 of
\citet{dg16a}, which shows a similar trend for bar size relative to near-IR
galaxy size as a function of stellar mass.\label{fig:barsize-R25}}

\end{figure}

The distributions of GZ2 and \sfourg{} bar sizes appear to converge
again for much lower stellar masses (e.g., $\logmstar \la 9$), though
the lower boundary for \sfourg{} sizes remains clearly below that for
the GZ2 galaxies. This behavior might reflect the increase in
\textit{relative} bar size (bar semi-major axis divided by galaxy
optical radius) at lower masses, as shown in
Figure~\ref{fig:barsize-R25}.

\section{Simulating SDSS Bar Observations}\label{sec:sims} 

As shown in the previous section, the distribution of bar sizes in the
local universe suggests that a significant number of bars could be
missed in SDSS-based surveys, for the simple reason that they are too
small in angular size to be reliably detected. 
From Figure~\ref{fig:barsize-vs-all-with-fwhm}, we can deduce that
SDSS-based studies will preferentially miss bars in lower-mass, bluer,
and more gas-rich galaxies, because those are the galaxies with smaller bars;
Figure~\ref{fig:barsizes-H11-vs-S4G} suggests that \textit{detected}
bars in GZ2 studies are indeed biased toward larger sizes relative to
local bars, and that this bias increases toward lower galaxy masses.

Can we test this idea more rigorously? The \sfourg-based subsamples I
use are defined to be volume- and mass-limited, which makes them
potentially a good match to many of the SDSS-based
samples.\footnote{Somewhat less so at lower masses, as the SDSS-based
studies are really absolute-magnitude-limited, so that stellar-mass incompleteness
starts to become a problem for $\logmstar < 10$ or so.} This means that it is
possible to use the \sfourg{} galaxies -- and their measured bar sizes
-- as a parent sample for simulating how local galaxies would be
classified if they were observed with typical SDSS resolution at typical
SDSS redshifts. Since the local subsamples are $\sim 97$\% complete in
terms of gas mass fraction, they are also suitable for examining simulated
bar fractions as a function of \fgas{} (although the incompleteness of
the actual SDSS studies in terms of \fgas{} makes comparison more
difficult).

The basic idea is simple: a mock sample of $N$ galaxies (where $N$ is
similar in size to typical SDSS samples) is generated by sampling with
replacement from an \sfourg-based subsample and assigning each galaxy a
random redshift drawn from the mock sample's redshift range, using
uniform volume weighting. Since the relevant SDSS-based studies
generally only have galaxies with $\logmstar \ga 9$--9.5
\citep[e.g.,][]{barazza08,masters12,cheung13,oh12,cervantes-sodi17}, the
best \sfourg-based subsample to use is probably Sample~2m: $\logmstar
\geq 9$, $D \leq 30$ Mpc. (For the gas-mass-ratio comparison, I
limit the starting simulation sample to $\logmstar \geq 9.5$.) If the
original \sfourg{} galaxy is unbarred, then the sampled galaxy is
assumed to be correctly identified as such. If the original galaxy
\textit{does} have a bar, then the orientation of the sampled galaxy's
bar within its disc is chosen randomly from $(0,90\degr)$ relative to
the line of nodes, the galaxy inclination $i$ is chosen randomly from
$(0,60\degr$), weighted by $\cos i$, and the observed bar size is
computed by projecting the intrinsic \sfourg{} bar semi-major axis
according to the chosen orientation. The galaxy is then ``observed'' to
determine if its bar is detected, with the projected bar size in kpc
converted to an angular size given the assigned redshift. If this is
larger than the assumed angular-size limit of $2 \times \textrm{FWHM}$,
then the galaxy is classified as ``barred''; otherwise, it is classified
as ``unbarred''.  The detected bar frequencies are then calculated for
each stellar-mass or gas-mass-fraction bin. The process is repeated 200
times, with the median and 68\% confidence intervals calculated from the
results.

A sharp cutoff based solely on bar size, such as I simulate here, is
obviously not terribly realistic. Bar \textit{strength} (however
defined) is probably important at some level, despite the evidence that
the SDSS-based studies are not preferentially selecting strong bars
(Section~\ref{sec:strong-bars}): bars which are, for example,
especially elliptical and high-contrast may still be detectable even if
their projected size is slightly below the resolution limit, while
rounder, lower-contrast bars may be missed even if their projected size
is slightly above the limit, even though both will be strongly affected
by convolution with the seeing. Another complication is that the actual
seeing for the SDSS images is variable; some images may have unusually
good seeing, making smaller bars more detectable, while other images
with worse-than-typical seeing will make detecting larger bars more
difficult. A further potential issue is that the \sfourg{} sample --
despite containing galaxies from the Virgo and Fornax Clusters -- does
not have any galaxies from very massive clusters, and so the
environmental match to the larger, SDSS-based samples is not perfect.
Finally, the \sfourg{} subsample I use excludes S0 galaxies, although
some of the SDSS-based studies exclude ``early-type'' galaxies
themselves.

These issues aside, I believe it is still useful to investigate how well
the SDSS-based bar-fraction trends can be reproduced -- even if only
qualitatively -- by this simple resolution-effects model. The next two
subsections explore this for the cases of stellar mass and gas mass
ratio.

\subsection{Simulated Bar Fraction and Stellar Mass} 

\begin{figure*}
\begin{center}
\hspace*{-1.5mm}\includegraphics[scale=0.86]{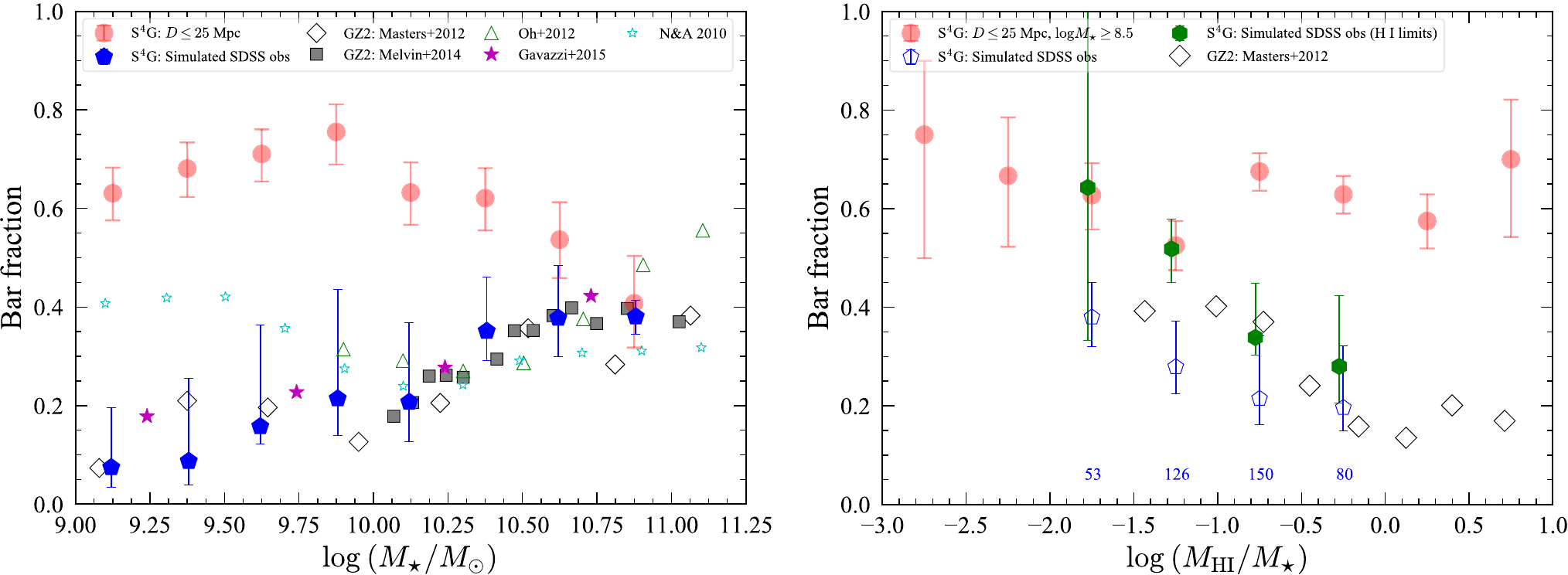}
\end{center}

\caption{Observed and simulated bar frequencies as a function of stellar
mass and \fgas. In both panels, observed bar fractions for \sfourg{}
galaxies are shown with filled red circles (see
Figure~\ref{fig:fbar-vs-all}). Left: bar fractions as function of
stellar mass, with SDSS-based fractions from \citet{masters12} and
\citet{melvin14} in grey, from \citet{oh12} with open green triangles,
from \citet{gavazzi15} with filled magenta stars, and from \citet{nair10b} with open
cyan stars. Filled blue pentagons
show median bar fractions (with 68\% confidence intervals) for simulated
observations of \sfourg{} galaxies at typical SDSS-survey redshifts ($z
= 0.01$--0.05), assuming that bars with (projected) semi-major axes
smaller than twice the typical FWHM of SDSS images are not detected.
Right: same, but now showing bar fractions as function of \fgas,
including GZ2 fractions from \citet{masters12}. Open blue pentagons are
from simulations similar to those in the left panel; green hexagons --
slightly offset for clarity -- are from simulations that include a
simplistic \hi{} selection function based on \citet{haynes11}. The
blue numbers along the bottom of the plot are the number of \sfourg{}
galaxies in each bin in the $\logmstar \geq 9.5$ parent sample used for the simulation.
\label{fig:fbar-sim}}

\end{figure*}

The left-hand panel of Figure~\ref{fig:fbar-sim} shows the median trend
for \fbar{} versus \logmstar{} from 200 bootstrapped mock samples, each
with $N = 10,000$ galaxies. The basic trend seen in GZ2
\citep{masters12,melvin14} and at least some other SDSS-based studies
\citep[e.g.,][]{oh12,gavazzi15,cervantes-sodi17} is clearly reproduced:
\fbar{} is very low for galaxies with $\logmstar < 10$, and then
increases steeply for higher masses.

The relation seen by \citet{nair10b} is not as well reproduced --
specifically, the increase in \fbar{} to \textit{lower} stellar masses
($\logmstar < 10$) that Nair \& Abraham reported is not apparent,
although GZ2 and other SDSS studies did not report much of a trend
in this sense, either.
In this case, difference in how bar detection works in the different
studies may be important. In particular, it could be that the Nair \&
Abraham study is more sensitive to both the \textit{relative} size of
bars and the absence of bulges in low-mass galaxies. The first factor is
illustrated by Figure~\ref{fig:barsize-R25} \citep[see also Fig.~20 of
][]{dg16a}, which shows that bar size relative to optical disc size
actually \textit{increases} to lower stellar masses for $\logmstar \la
10$. It may thus be easier in some cases to identify bars in lower-mass
galaxies simply because they span a larger part of the visible disc. The
\textit{minimum} in relative bar size happens at $\logmstar \approx
10.2$ -- which is where the \citet{nair10b} bar frequency has \textit{its}
minimum. This might also explain why \citet{cervantes-sodi17} reports a
sharp upturn in \fbar{} for his very lowest-mass bin (his Fig.~1), along
with the weak upturns in \fbar{} for very blue and low-mass galaxies in
GZ2 \citep[e.g.,][]{masters11,masters12}.

A second possible reason for the high \fbar{} values seen for low masses
(and blue galaxies) by \citet{nair10b} is that their sample is
\textit{magnitude-limited} ($g < 16$), unlike e.g.\ the GZ2 studies, and
unlike the volume-limited subsamples of \sfourg{} constructed for this
paper. A magnitude limit means that less luminous galaxies (which tend
to be lower in mass and bluer) are preferentially found at smaller
redshifts, as can be seen in Fig.~20 of \citet{nair10a}: for example,
galaxies with $\logmstar = 8.0$--9.5 are mostly found at $z \approx
0.01$--0.02, while galaxies with $\logmstar = 10.8$--11.2 are mostly
found at $z \approx 0.04$--0.06. This means that a typical low-mass
galaxy will be observed with several times better spatial resolution
than a typical high-mass galaxy in their sample. It is conceivable that
the upturn in \fbar{} toward lower stellar masses in Fig.~1 of
\citet{nair10b} is at least partly due to this effect -- especially
since the \sfourg{} samples show that galaxies with $\logmstar \la 10.1$
have bar sizes that depend rather weakly on stellar mass
(Figure~\ref{fig:barsize-with-fit}).

Unfortunately, I cannot properly test this last hypothesis using the
\sfourg{} sample, because the $g = 16$ limit of \citet{nair10a} is
almost a full magnitude fainter than the \sfourg{} parent-sample
magnitude limit of $B = 15.5$. This means that the Nair \& Abraham
sample has many galaxies fainter than are found in \sfourg{}, making it
difficult to compare the samples.

Leaving aside the caveats regarding the \citet{nair10b} study, I
conclude that the interaction of the bar-size--stellar-mass correlation
and low spatial resolution explains most, if not all, of the apparent
correlation between \fbar{} and stellar mass reported by the majority of
SDSS-based studies.

\subsection{Simulated Bar Fraction and Gas Mass Ratio} 

The right-hand panel of Figure~\ref{fig:fbar-sim} shows simulated SDSS
observations of \fbar{} as a function of \fgas. Results from simulations
using the same approach as for the \logmstar{} comparison -- except now
using a sample size of 2000 as a better match to that used by
\citet{masters12}, and with stellar masses restricted to $\logmstar \geq
9.5$ -- are shown using hollow blue pentagons. As expected, the observed
bar fraction decreases to higher gas mass ratios. This qualitatively
agrees with the results of \citet{masters12} and
\citet{cervantes-sodi17}. The \textit{quantitative} match is not as good
as the agreement between the simulated and reported \fbar--\logmstar{}
results (left-hand panel of the figure). This is at least partly because
the SDSS \fgas{} results include complicated selection effects based on
the varying detection efficiency of the ALFALFA survey as a function of
gas content and redshift \cite[see the discussion in][]{masters12}.

Inspection of Fig.~2 of \citet{masters12} suggests that two related
biases may be relevant. Most of their gas-rich galaxies (e.g., $\logfgas
= 0$--0.5) are low-mass -- and thus have smaller intrinsic bar sizes
-- \textit{and} are mostly at higher redshifts (the majority between $z
\sim 0.03$ and 0.05). In contrast, most of their gas-\textit{poor}
galaxies (e.g., $\logfgas = -1.5$ to $-1.0$) are high-mass -- and thus
have \textit{larger} intrinsic bar sizes -- and at the same time are at
\textit{lower} redshifts (mostly between $z \sim 0.015$ and 0.035). This
is of course a natural result for the combination of a flux-limited \hi{} survey 
and the strong anticorrelation between gas content and stellar mass. So
gas-rich galaxies in SDSS + ALFALFA samples will tend to have small
intrinsic bar sizes \textit{and} be further away, which will lower the
bar-detection efficiency, while gas-poor galaxies will have
intrinsically larger bars and be closer, which will \textit{increase}
the bar-detection efficiency. This effect is a plausible explanation for
the fact that \citet{masters12} and \citet{cervantes-sodi17} find
\fbar{} to be anticorrelated with gas mass ratio at fixed stellar mass
(or, equivalently, \fbar{} increases with \hi{} deficiency, which is
computed as a function of stellar mass).

As an attempt to include the effects of the \hi{} detection bias of the
SDSS studies, I generated additional simulated surveys using the
suggested 50\% detection limit for ALFALFA from Eqns.~6 and 7 of
\citet{haynes11}, with the HyperLeda \texttt{vmaxg} parameter standing
in for $W_{50}$ (i.e., $W_{50} = 2$ \texttt{vmaxg}). For each potential
\sfourg{} galaxy chosen for the bootstrapped sample, I computed its
hypothetical \hi{} flux (using the randomly chosen redshift and the
galaxy's actual \hi{} flux from HyperLeda) and the ALFALFA limiting
flux. If the hypothetical flux at redshift $z$ was brighter than the
limiting flux, then the galaxy was accepted into the bootstrapped
sample; if its \hi{} flux was too faint, it was rejected and a new
galaxy selected from the parent sample, with the process repeated until
a galaxy was accepted. The results of this ``\hi--limited'' simulation
are shown using green hexagons in the right panel of
Figure~\ref{fig:fbar-sim}. These observed bar fractions are higher than for the
standard simulation (hollow blue pentagons), because the
\hi--selection cutoff tends to exclude more distant galaxies, where bars
are harder to detect. The overall pattern of increasing bar fraction to
lower \fgas{} is even stronger, and the predicted \fbar{} values tend to
overlap somewhat better with the GZ2 values from \citet{masters12}.

This suggests that we can explain the discrepancy between the absence of
a dependence of bar fraction on gas fraction in \sfourg{} galaxies and
the apparent strong dependence seen in SDSS studies, at least to first
order, via a  combination of the dependence of bar size on stellar mass -- and
thus indirectly on \fgas{} -- \textit{and} the effects of strong \hi{} flux
limits.

\subsection{Bar Fraction and Colour} 

\citet{masters11,masters12} and \citet{lee12a} reported strong trends
between bar fraction and galaxy optical colour, in the sense that redder
galaxies are much more likely to be barred. As noted in
Section~\ref{sec:fbar-color-s4g}, the local (\sfourg) galaxies do not
show any evidence for this; if anything, the local trend suggests a very
slight (though statistically not significant) \textit{decrease} in
\fbar{} for redder galaxies; a similar trend was seen by
\citet{barazza08}.

Unfortunately, the incompleteness of colour data for \sfourg{} galaxies
(see Appendix~\ref{sec:colors}) makes it difficult to construct
simulated SDSS observations as I have done for the stellar-mass and
\fgas{} cases, particularly since the incompleteness is worse for bluer
galaxies.

Nonetheless, given that galaxy colour \textit{does} correlate with
stellar mass, it is very likely that the reported \fbar--colour
correlations from SDSS-based studies are mostly if not entirely side
effects of differential bar detection: since bluer galaxies tend to be
lower mass, they also tend to have smaller bars, which will be harder to
detect (middle panel of Figure 7). The fact that the mean resolution of
\citet{barazza08} was about twice as good as that of the GZ2 and
\citet{lee12a} studies, combined with a possible higher bar-detection
efficiency, likely explains why the Barazza et al.\ colour trend is
similar to the \sfourg{} trend.

\section{Discussion}\label{sec:discuss} 

\subsection{The Primacy of Stellar Mass} 

Section~\ref{sec:fbar-mass-s4g} suggests that bar presence depends quite
strongly on stellar mass, with little or no dependence on galaxy colour
or (neutral) gas content. 
We can still ask whether -- once we account for the main stellar-mass
dependence -- there might not still be some \textit{residual} dependences of
bar frequency on colour or gas content. Since the quadratic logistic
regression is such a good match to the bar frequency as a function of
stellar mass, I test for possible residual correlations by performing
logistic regressions using both stellar mass and one or both of the
other parameters as independent variables. This is in essence a
variation of Equation~\ref{eq:logistic1}, with the bar frequency
described as
\begin{equation}
\fbar(x) \; = \; \frac{1}{1 + e^{-(\alpha + \beta_{1} m + \beta_{2} m^{2} + \beta_{3} x)}} ,
\end{equation}
where $m = \logmstar$ and $x =$ either \gmr{} or \logfgas.

Table~\ref{tab:logistic2} shows the resulting best-fit parameters for the
fit using stellar mass and colour and the fit using stellar mass and \fgas{}.
Each fit is done to two subsets of data: galaxies in Sample 1m with \fgas{}
values (556 galaxies) and galaxies in Sample 1m with \gmr{}
values (319 galaxies). Akaike Information Criteria (AIC) values from the fits
can be used to evaluate the relative goodness of different fits, but only
within the individual subsamples. In both subsamples, the lowest AIC values --
and thus the preferred fits -- are for the quadratic dependence on \logmstar{} alone.
Adding either gas mass ratio or colour to the fits results in higher AIC values --
significantly so ($\Delta$AIC = 6.6) for the large, \fgas-based sample.
This is a reasonably clear indication that the bar frequency does not show
any significant dependence on either gas mass ratio or \gmr{} colour, once the
strong dependence on stellar mass is accounted for.

\begin{table}
\caption{Logistic Regression Results: Multiple Variables}
\label{tab:logistic2}
\begin{tabular}{@{}lrrrr}
\hline
Variable             & $\alpha$   & $\beta$   & $P_{\beta}$ &  AIC \\
(1)                  & (2)        & (3)       & (4)         & (5) \\
\hline
\multicolumn{5}{c}{Subsample: 556 galaxies with \fgas{} values} \\
$\log \Mstar$        &   $-6.05$  & 0.69     & 0.0012       & 799.6 \\[2mm]

$\log \Mstar$        &  $-84.29$  & 17.55    & 0.00016      & 762.1 \\
$(\log \Mstar)^{2}$  &  ---       & $-0.90$  & 0.00019      & --- \\[2mm]

$\log \Mstar$        &  $-84.51$  & 17.51    & 0.00020      & 768.7 \\
$(\log \Mstar)^{2}$  &  ---       & $-0.90$  & 0.00024      & --- \\
\logfgas{}           &  ---       & 0.21     & 0.55         & --- \\

\hline
\multicolumn{5}{c}{Subsample: 319 galaxies with \gmr{} values} \\
$\log \Mstar$        &   $-6.05$  & 0.69     & 0.0012       & 439.2 \\[2mm]

$\log \Mstar$        &  $-75.16$  & 15.35    & 0.037        & 433.5 \\
$(\log \Mstar)^{2}$  &  ---       & $-0.78$  & 0.037        & --- \\[2mm]

$\log \Mstar$        &  $-67.25$  & 13.62    & 0.071        & 434.5 \\
$(\log \Mstar)^{2}$  &  ---       & $-0.68$  & 0.077        & --- \\
\gmr{}               &  ---       & $-1.47$  & 0.21         & --- \\

\end{tabular}

\medskip

Results of multivariable logistic regression for bar frequency \fbar{}
in Sample~1m (\sfourg{} spirals with $D \leq 25$ Mpc and $\logmstar >
8.5$) as a function of log of stellar mass and either \gmr{} colour or
gas mass fraction \fgas. Two subsamples are considered: the first is
all galaxies in Sample~1m with \fgas{} values, while the second is
all galaxies with \gmr{} colours. (1) Galaxy parameter used in fit. (2)
Intercept value for fit. (3) Slope for fit. (4) $P$-value for slope. (5)
Akaike Information Criterion value for fit; values should be compared only within
a given subsample, with lower values indicating better fits.

\end{table}

\subsection{Implications for High-Redshift Bar Studies} 

Could differential resolution affect high-$z$ bar studies? Studies using
\textit{HST} data of galaxies at moderate redshifts (e.g. $z \sim
0.4$--1.0) sometimes make the same sort of claims that they are 
incapable of missing ``normal'' bars that some SDSS-based studies have
made. Whether this is true depends partly on resolution and partly on
what the (unknown) distribution of bar sizes at different redshifts
actually is. The spatial resolution of \textit{HST} images for
moderate-redshift galaxies is \textit{better} than typical $z \sim 0.05$
SDSS images: for example, even at $z = 1$, an $I$-band \textit{HST}
image will have a FWHM of $\sim 0.8$ kpc, compared with 1.25 kpc for a
typical SDSS image at $z = 0.05$. Clearly, fewer bars should be missed
by \textit{HST} surveys -- \textit{if} bars have the same size
distribution at high redshift as they do locally.

\begin{figure}
\begin{center}
\hspace*{-3mm}\includegraphics[scale=0.46]{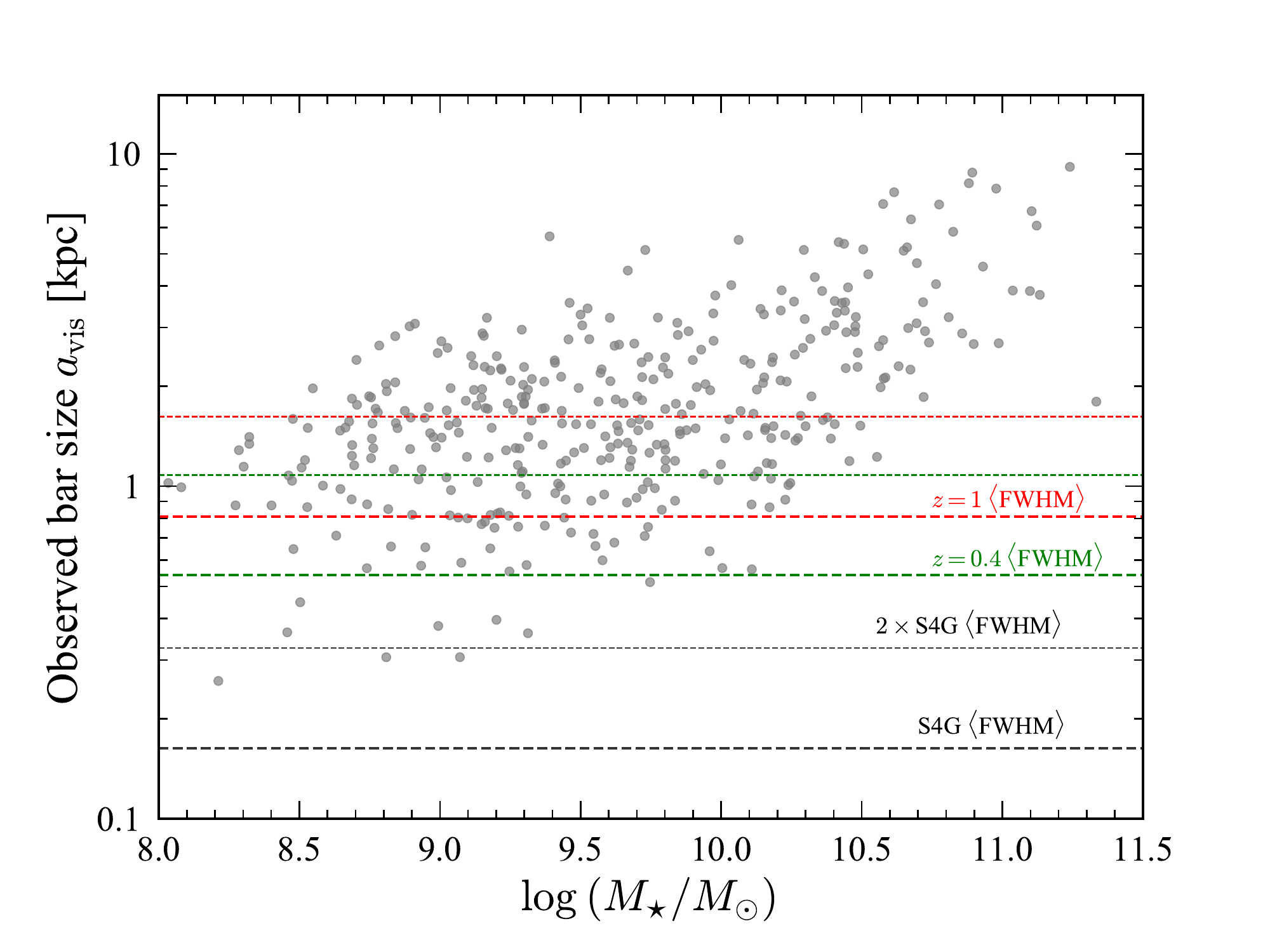}
\end{center}

\caption{As for Figure~\ref{fig:barsize-vs-all-with-fwhm} (upper panel),
but now showing potential resolution limits at intermediate to high
redshifts. Assuming a typical F814W FWHM of $\sim 0.09\arcsec$, the
expected resolution limits (FWHM) at $z = 0.4$ and 1.0 are shown with
thick dashed green and red lines, respectively. Thin dashed lines show
the $2 \times \mathrm{FWHM}$ limit for each case.
\label{fig:barsize-vs-mstar-highz}}

\end{figure}

Figure~\ref{fig:barsize-vs-mstar-highz} compares potential high-redshift
resolution limits with the \sfourg{} bar-size distribution: I show
linear FWHM for two different redshifts ($z = 0.4$ and 1.0, bracketing
the range usually used in \textit{HST} bar studies), assuming an
$I$-band angular PSF FWHM of 0.09\arcsec{} (e.g., for F814W or F850LP
filters used with ACS-WFC or WFC3-UVIS). If bars can be reliably
identified in \textit{HST} images down to $a \sim$ FWHM, then relatively
few bars should be missed, at least for $\logmstar \ga 10$. But if a
more conservative limit of $2 \times {\rm FWHM}$ is used -- as would
seem wise for GZ2-style studies such as \citet{melvin14} and
\citet{simmons14} -- then we might expect lower bar fractions at higher
redshifts purely from resolution effects, and this could apply to
stellar masses as high as $\sim 10^{10.5} \Msun$.

There are actually two things to worry about in terms of resolution for
\textit{HST} bar studies. The first is the high-redshift equivalent of
what happens for the low-redshift SDSS studies: the fact that a
resolution limit near or above the size of the smallest bars, combined
with a dependence of bar size on galaxy mass, can create -- or
exaggerate -- a trend of \fbar{} with stellar mass. The second is that
comparing bar fractions at different redshifts means comparing different
resolutions, for the simple reason that higher-redshfit samples are
observed with lower spatial resolution. As
Figure~\ref{fig:barsize-vs-mstar-highz} shows, going from $z = 0.4$ to
$z = 1.0$ will lower the detected bar fraction, creating an apparent
evolution in the \fbar{} with redshift qualitatively similar to
what has actually been reported.

One way to address the second confounding factor -- the change in
resolution with redshift -- would be to convolve images of
lower-redshift galaxies to the worst-case spatial resolution of
the high-redshift limit of a survey, so that all galaxies are observed
with the same spatial resolution. Unfortunately, this seems to be
something that is never attempted.

The preceding argument is based on the undoubtedly unrealistic
assumption that bars at high redshift having the same distribution of
sizes that they do in the local universe. Most theoretical studies and
models suggest that bars should grow in size over time
\citep[e.g.,][]{debattista00, martinez-valpuesta06,algorry16}, generally
by factors of $\sim 50$--100\%. But this makes the resolution problem
\textit{worse}, because bars at higher redshifts would then be
\textit{smaller} and thus even harder to detect. 

\begin{figure}
\begin{center}
\hspace*{-2mm}\includegraphics[scale=0.69]{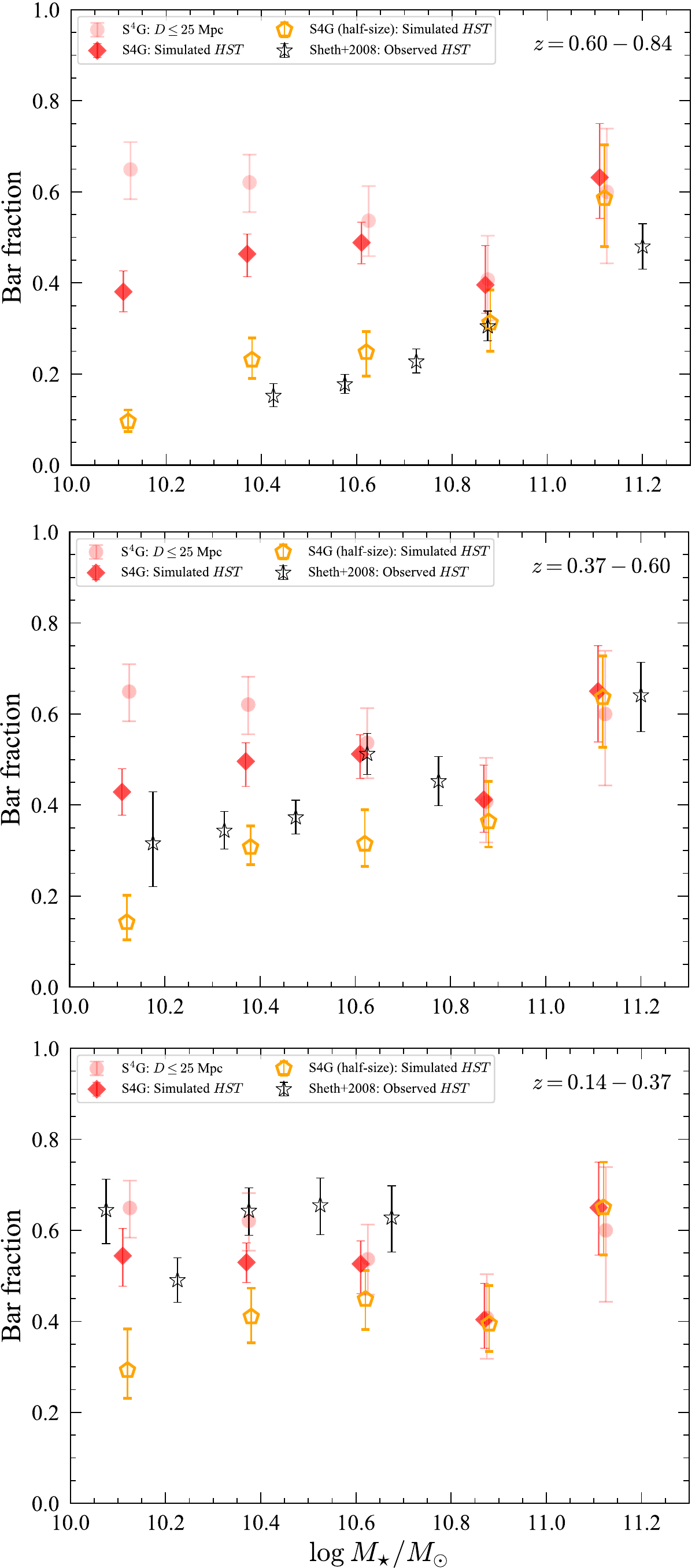}
\end{center}

\caption{As for the left panel of Figure~\ref{fig:fbar-sim}, but now
showing simulated high-redshift observations with \textit{HST}. Red
circles show the local observed \sfourg{} bar fractions; red diamonds
show simulated \textit{HST} observations of \sfourg{} spirals assuming
that bars with projected semi-major axes $< 2 \times \textrm{FWHM}$ go undetected
(using FWHM $= 0.09\arcsec$), while orange pentagons show the same thing
but also assuming that bar sizes were only half their present value.
Black stars show observed \textit{HST} total bar fractions from
\citet{sheth08}. Top: Simulations and data for $z = 0.60$--0.84. Middle:
$z = 0.37$--0.60. Bottom: $z = 0.14$--0.37.
\label{fig:fbar-vs-mstar-highz-sim}}

\end{figure}

A simplistic demonstration of the possible effects is shown in
Figure~\ref{fig:fbar-vs-mstar-highz-sim}, which repeats the simulated
observations approach of Section~\ref{sec:sims}, but places all
resampled \sfourg{} galaxies at three different redshift ranges,
corresponding to the ranges used in the study of \textit{HST} COSMOS
data by \citet{sheth08}, and observes them with a typical \textit{HST}
$I$-band FWHM of 0.09\arcsec. Red diamonds show the results using the
\sfourg{} spiral sample as is, while the orange pentagons show the results
when their bar sizes are set to half their actual values. The black stars
are the observed \fbar{} values from \citet{sheth08}.\footnote{These
points are plotted at the center of their respective bins, in contrast
to the original plot in Fig.~3 of \citet{sheth08}, where they were
plotted at the left side of their bins.}

If bars do \textit{not} grow significantly in size, so that they were
already as large at higher redshifts as they are now, then the
simulations overpredict bar frequencies for all galaxies in the high
redshift regime ($z = 0.60$--0.84, top panel of
Figure~\ref{fig:fbar-vs-mstar-highz-sim}), and they overpredict bar
frequencies for lower-mass ($\logmstar \la 10.5$) galaxies in the medium
redshift regime ($z = 0.037$--0.60, middle panel)). For the lowest
redshift regime ($z = 0.14$--0.37, bottom panel), the observed
frequencies are generally consistent with the local values -- and
possibly more consistent with observed true local values than the
prediction for that redshift bin; this might be an indication that the
$2 \times \mathrm{FWHM}$ detection limit I assume underestimate the
bar-detection efficiency of the methods used in \citet{sheth08}.

If, on the other hand, bars at higher redshifts were only half their
present size, then the simulations are a surprisingly good match for the
high-redshift sample. In particular, the trend of increasing bar
fraction for increasing stellar mass is fully reproduced, and the
predicted frequencies are only marginally higher than the observed
\textit{HST} frequencies. For the intermediate redshifts, the
simulations with smaller bar sizes slightly \textit{underpredict}
the observed frequencies, and they clearly underpredict the observed
frequencies in the lowest-redshift bin (bottom panel); this suggests
that at those redshifts bars were already more than half their present
size.

Taken together, these result suggest that a combination of significant
growth in bar size \textit{and} redshift-dependent resolution effects
can in principle explain most, though perhaps not all, of the observed
changes in bar frequency reported by \citet{sheth08}. In particular, the
simulations do a good job of reproducing the trends in bar frequency as
a function of stellar mass -- i.e., increasing with stellar mass over
the range of $\logmstar \sim 10.2$--11 -- which are similar to those
seen in most SDSS-based samples, and which differ so dramatically
from the \sfourg{} trend (Figure~\ref{fig:fbar-vs-all}).

A model where all bars were already in place at $z\sim 0.8$ but had
only half their size, so that the observed redshift evolution is not due
to any actual changes in bar \textit{frequency}, is probably not realistic. The
truth is almost certainly somewhere in between: that is, the apparent
evolution in bar fraction with redshift reported by \textit{HST}
studies is probably real, but has likely been exaggerated by the
combination of resolution effects and changes in bar sizes over cosmic
time. Before we can say for certain how the bar fraction evolves with
redshift -- and with stellar mass -- we need to be able to disentangle
true evolution in bar frequencies from the evolution of resolution
effects, and the evolution of bar sizes.

\subsection{Implications for Models of Bar Formation and Growth} 

Theoretical studies of bar formation and evolution have suggested that
bar formation in gas-rich discs is suppressed or at least delayed, and
that once a bar forms its growth can also be slowed when the disc is
gas-rich \citep[e.g.,][]{berentzen98,debattista06,athanassoula13}.
Alternately, too much gas accretion might weaken or even destroy
pre-existing bars \citep{bournaud02,bournaud05c}. The SDSS-based results
of \citet{masters12}, \citet{cheung13}, and \citet{cervantes-sodi17}
have been interpreted to support at least some of this, since they found lower bar
fractions in galaxies with more gas (or in galaxies with higher
star-formation rates, which can be interpreted as a proxy for gas
content). 

However, the fact that local galaxies show an essentially
\textit{constant} bar fraction as a function of (atomic) gas mass
fraction (bottom panels of Figure~\ref{fig:fbar-vs-all}) runs counter to
much of the preceding. It clearly disagrees with the predictions of
\citet{bournaud02} and \citet{bournaud05c} that high gas content should
rapidly destroy bars, although this is perhaps not surprising, as most
recent simulations have found that bars are rarely if ever destroyed by
high gas content \citep[see., e.g., the discussion
in][]{athanassoula13}. More generally, the fact that the majority of
local spirals have $\fgas \ga 0.1$
(Figure~\ref{fig:gmr-and-fgas-vs-mstar}) and that the observed bar
fraction is roughly constant from values of $\fgas \sim 0.01$ to 1
suggests that high gas content cannot be very effective in suppressing
bar formation or growth over long time periods -- unless any gas-driven
delay is short enough that essentially all galaxies have already passed
through it and formed their bars.

Only for very high gas mass fractions is there some evidence for an
effect: as can be seen in the lower-right panel of
Figure~\ref{fig:fbar-vs-all}, the fraction of \textit{strong} bars
declines rather steeply for $\fgas \ga 1$, while the fraction of
\textit{weak} bars goes up. This gas-mass-fraction level is higher than
that explored in most bar simulations.

\section{Summary} 

In order to investigate how the local frequency of bars in spiral
galaxies depends on galaxy mass and other properties, I constructed
distance- and mass-limited subsamples of low-inclination ($i \leq
65\degr$) spiral galaxies from the \textit{Spitzer} Survey
of Stellar Structure in Galaxies (\sfourg), using bar identifications
and measurements from \citet{herrera-endoqui15} and stellar masses from
\citet{munoz-mateos15}. I included colour and neutral-hydrogen
measurements from HyperLeda and computed $V/V_{\rm max}$ weights to
correct for the angular-diameter limit of \sfourg; for the colours, I
also computed $B_{\rm tc}$-based weights to correct for colour
incompleteness in the HyperLeda data. The resulting subsamples include 851
galaxies with $D \leq 30$ Mpc and 659 with $D \leq 25$ Mpc; the latter
is the main subsample analyzed in this paper. Applying stellar-mass
limits yields 638 galaxies with $\logmstar \geq 9$ and $D \leq 30$ Mpc,
or 574 galaxies with $\logmstar \geq 8.5$ and $D \leq 25$ Mpc.

The main findings of this analysis are the strong dependence of bar
frequency on stellar mass and the \textit{absence} of any clear
dependence on galaxy colour or gas content. More specifically, the bar
frequency in \sfourg{} galaxies peaks at $\logmstar \approx 9.7$ with a
value of $\fbar = 0.70$ and declines to both lower and higher masses. On
the other hand, \fbar{} is roughly \textit{constant} as a function of
\gmr{} colour (or perhaps declining to redder colours, though this trend
is not statistically significant) over the range $\gmr \approx
0.1$--0.8, and is also roughly independent of \hi{} gas mass fraction
($\fgas = \MHI / \Mstar$).

These \fbar{} trends differ quite strongly from most studies using large
SDSS-based samples (typically at $z = 0.01$--0.5), which tend to find
that \fbar{} increases strongly toward \textit{higher} masses (peaking
at $\logmstar \sim 11$), redder colours, and lower values of \fgas. The
SDSS-based \fbar{} values are almost always significantly lower than the
\sfourg{} galaxies, except for the highest stellar masses and reddest
colours.

I argue that this discrepancy between very local (\sfourg) and SDSS-based
bar fractions can be explained primarily by a combination of two factors:
\begin{enumerate}

\item Differences in effective spatial resolution due to differences in
galaxy distances. Galaxies that are further away have bars that are
harder to detect, given the finite angular resolution of images. The
effective spatial resolution of \sfourg{} is $\sim 0.17$ kpc at the mean
distance (17 Mpc) of the $D \leq 25$ Mpc subsample, assuming an angular
FWHM of $\approx 2\arcsec$. The equivalent for typical SDSS-based
studies (mean redshift $\sim 0.045$, typical FWHM $\sim 1.4\arcsec$) is
$\sim 1.3$ kpc.  A plausible angular-size limit for bar detection
is $a_{\rm obs} \sim 2 \times \mathrm{FWHM}$, which makes the mean linear-size
limits $\sim 0.33$ kpc for \sfourg{} and $\sim 2.5$ kpc for typical
SDSS-based surveys.

\item The dependence of bar size on galaxy stellar mass (and, less
directly, on colour and gas content). Bar size is correlated with
stellar mass (in a bimodal fashion, with a steeper relation for
$\logmstar \ga 10.2$); bar sizes also exhibit significant scatter
around the mean value for a given stellar mass.
\end{enumerate}
The result is that bars in high-mass (red, gas-poor) galaxies are easier to
find because they are larger. The effective resolution of the
\textit{Spitzer} images of \sfourg{} $D \leq 25$ Mpc subsample is good
enough so that almost all bars have observed semi-major axes -- including
projection effects -- larger than $2 \times \mathrm{FWHM}$ and can thus be
detected, even for quite low stellar masses. But for distances typical
of SDSS-based surveys, bar sizes start to drop below $2 \times \mathrm{FWHM}$ for
stellar masses $\logmstar \la 10.7$. The dispersion of bar sizes means
that significant numbers of bars can be missed even though the
\textit{mean} bar size might still be above the resolution limit.

Comparison of the observed sizes of bars from GZ2, as tabulated by
\citet{hoyle11}, with sizes for \sfourg{} galaxies supports this
general argument. GZ2 and \sfourg{} bars have similar sizes at very high masses
($\logmstar \sim 11$); but as one goes to lower masses, the average for
\sfourg{} bars falls below that of GZ2 bars, until by $\logmstar \sim
10.2$, the average \sfourg{} bar is smaller than $\sim 90$\% of GZ2
bars. This is a clear indication that GZ2 is systematically missing
smaller bars at lower masses, which will mean a lower apparent bar
fraction. (The bars that are being missed are not a separate population
of ``nuclear'' bars, but simply the lower end of the distribution of
normal galactic bars.)

An additional bias applies to SDSS-based studies using \fgas{}
\citep{masters12,cervantes-sodi17}, because only galaxies with enough
gas to be detected by ALFALFA were included in the samples. This means
that galaxies with high gas mass fractions tended to be observed out to
larger distances than galaxies with low gas mass fractions. Gas-rich
galaxies thus had, on average, larger distances and therefore lower
spatial resolution, making their bars harder to identify.

To see if the specific \fbar{} trends of typical SDSS-based studies could
be explained by these effects, I simulated SDSS observations by
constructing mock samples using sampling with replacement from \sfourg{}
galaxies with $D < 30$ Mpc, placing the galaxies at random (volume-weighted)
redshifts between 0.01 and 0.05; \fbar{} was then computed by assuming that
bars with observed sizes less than twice the typical SDSS FWHM were not
detected. This did an excellent job of reproducing the reported
\fbar-\logmstar{} trend from GZ2 studies, and at least partly reproduced
the reported \fbar-\logfgas{} trends of \citet{masters12} and
\citet{cervantes-sodi17}.

The absence of a clear trend between bar presence and gas content in local
galaxies casts doubt on models in which bar formation is suppressed by high gas
fractions in the disc, unless any such suppression is temporary only, and ended
significantly before the present epoch.

Finally, I note that differential detection effects due to bar-size
correlations and changing distances could also affect high-redshift
studies of the bar fraction, since even \textit{HST}'s resolution is not
good enough to prevent smaller bars from being missed, especially at
higher redshifts. The apparent decrease in \fbar{} reported for
high-redshift samples -- and the reported trends of \fbar{} increasing
to higher stellar masses over the range $\logmstar \sim 10$--11 at high
redshifts -- could also be influenced by secular changes in bar size:
since most theoretical studies argue that bars grow in length over time,
bars at higher redshifts would be smaller in size, and thus even harder
to detect. Simulations similar to those done for the SDSS-based studies
suggest that the combination of changing resolution and growth in bar
sizes could account for at least some -- though probably not all -- of
the reported evolution of bar fractions with redshift.

\section*{Acknowledgments} 

This study benefited greatly from conversations with, and
encouragement from, Victor Debattista.  I also thank Dave Wilman and
Ariel Sanchez for useful discussions and Martin Herrera-Endoqui for
answering questions about the \sfourg{} measurements and comments on an
earlier draft. Additional comments from Sim{\'o}n D{\'\i}az-Garc{\'\i}a,
Dimitri Gadotti, Karen Masters, Brooke Simmons, Sandor Kruk, Chris
Lintott, and David Trinkle are greatly appreciated. I also
thank the referee, Johan Knapen, for a number of helpful suggestions
which improved this manuscript.

Funding for the creation and distribution of the SDSS
Archive has been provided by the Alfred P. Sloan Foundation, the
Participating Institutions, the National Aeronautics and Space
Administration, the National Science Foundation, the U.S. Department of
Energy, the Japanese Monbukagakusho, and the Max Planck Society. The
SDSS Web site is \texttt{http://www.sdss.org/}.

The SDSS is managed by the Astrophysical Research Consortium (ARC) for
the Participating Institutions.  The Participating Institutions are
The University of Chicago, Fermilab, the Institute for Advanced Study,
the Japan Participation Group, The Johns Hopkins University, the
Korean Scientist Group, Los Alamos National Laboratory, the
Max-Planck-Institute for Astronomy (MPIA), the Max-Planck-Institute
for Astrophysics (MPA), New Mexico State University, University of
Pittsburgh, University of Portsmouth, Princeton University, the United
States Naval Observatory, and the University of Washington.

This research also made use of Astropy, a community-developed core Python
package for Astronomy \citep{astropy13}.

\bibliographystyle{mnras}

\appendix 

\section{Galaxy Colours}\label{sec:colors} 

\subsection{Correcting for Incompleteness in Galaxy Colours}\label{sec:color-incompleteness} 

\begin{figure}
\begin{center}
\hspace*{-3mm}\includegraphics[scale=0.47]{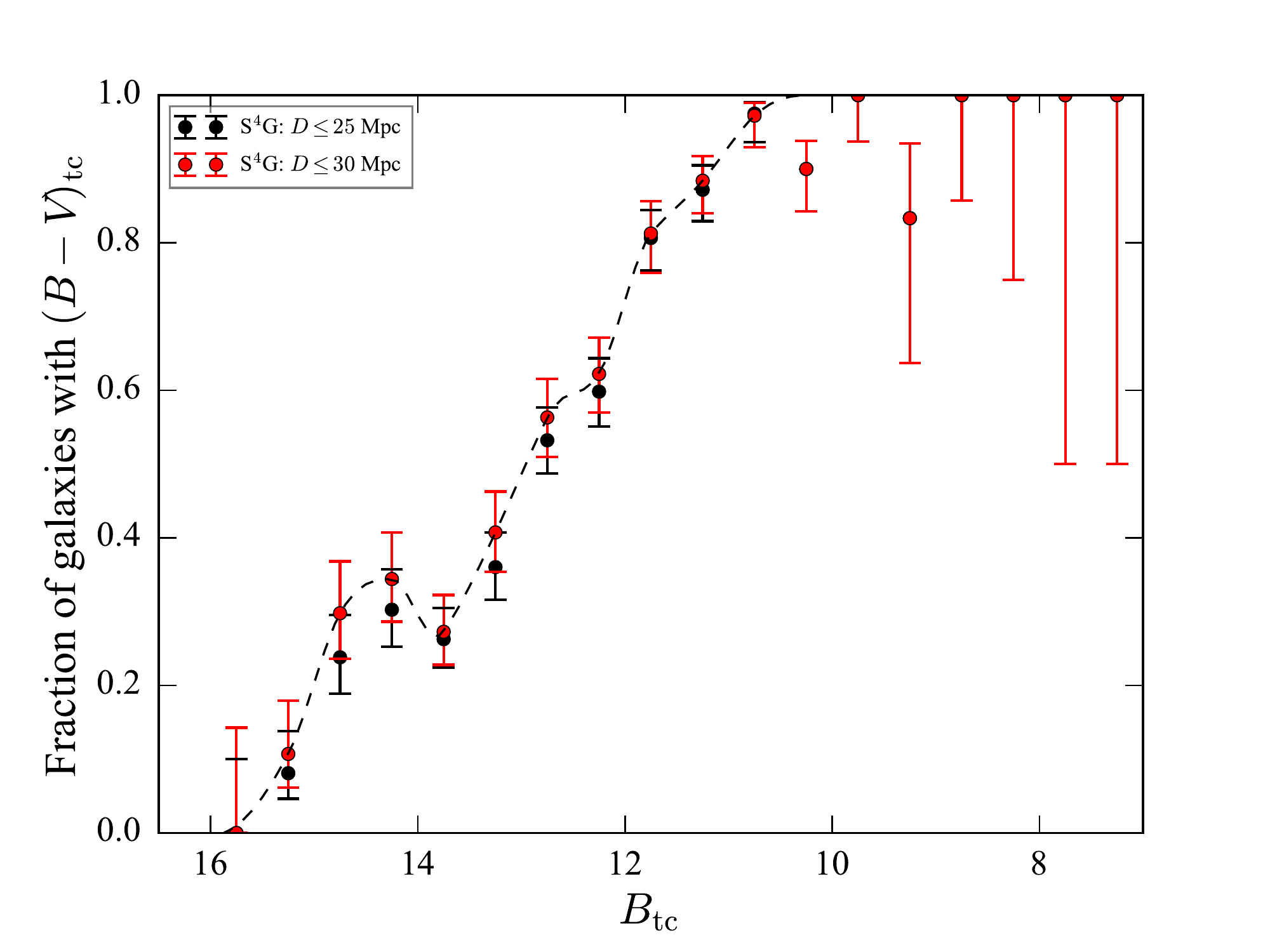}
\end{center}

\caption{Completeness of \bmvtc{} colours from HyperLeda for the
\sfourg{} sample as a function of apparent blue magnitude \btc{} for the
two distance-limited subsamples (black, red). The dashed black line
shows the spline interpolation (to the $D \leq 30$ Mpc subsample) used
to generate galaxy weights for colour-based
analyses.\label{fig:bmv-vs-btc}}

\end{figure}

The HyperLeda \bmvtc{} colours are the largest set of homogeneous,
whole-galaxy colours available for very nearby galaxies such as those in
the \sfourg{} sample. (SDSS colours are available for many more galaxies
in general, but the sky coverage is limited and very nearby galaxies are
prone to suffer from the ``shredding'' effect, so that they do not have
accurate SDSS magnitudes or colours -- see, e.g., \citealt{consolandi16a}).
Even so, the HyperLeda colours are highly incomplete: about half of the
galaxies in the \sfourg{} sample do not have such colours. What is
potentially worse is that the incompleteness is not random; as
Figure~\ref{fig:bmv-vs-btc} shows, it is a strong function of apparent
magnitude.

To correct for this incompleteness, I weight individual galaxies in
colour-based analyses by the inverse of their apparent-magnitude-based
colour completeness. To generate these weights, I model the observed
completeness of \bmvtc{} values as a function of \btc{} using a cubic
Akima spline interpolation,\footnote{Akima splines provide a smoother
interpolation, with less ringing between data points, than do a standard
cubic splines.} as provided in the Scipy \texttt{interpolate} module
(curve in Figure~\ref{fig:bmv-vs-btc}). To allow the interpolation to
asymptote smoothly to zero for faint \btc{} values (and thus avoid
potentially infinite weights for very faint galaxies), I added two
artificial points at $\btc = 16$ and 17. Since only $\sim 1\%$ of the
sample galaxies are fainter than $\btc = 15.5$, the effect is minimal.

In Section~\ref{sec:fbar-color-s4g}, I compute the \sfourg{} bar
fraction as a function of \gmr{} colour, since the latter makes for a
direct comparison with previously published SDSS bar studies
\citep[e.g.,][]{barazza08,masters11,masters12}. The \gmr{} colours are
based on the \bmvtc{} colours, using the empirical, galaxy-based
transformations of \citet{cook14b}, which work out (from their Table~3)
as $\gmr = 1.12 \, (\bmv) - 0.18$.

\subsection{Bar Fractions as Function of \bmv{}}\label{sec:fbar-vs-BmV} 

\begin{figure*}
\begin{center}
\hspace*{-5mm}\includegraphics[scale=0.87]{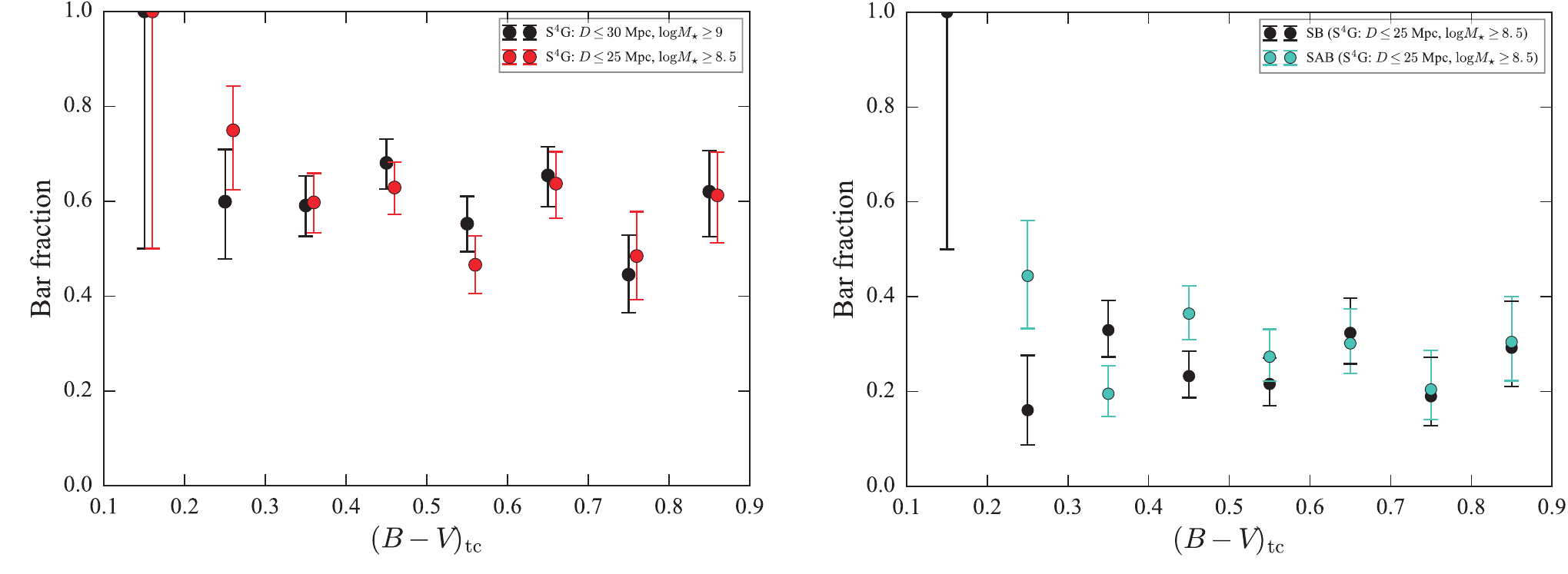}
\end{center}

\caption{Left: Fraction of disc galaxies with bars as a function of
total, extinction-corrected \bmv{} colour for \sfourg{} galaxies in
Samples~1m (black) and 2m (red). Right: Fractions of disc galaxies with
strong (black) and weak (cyan) bars as a function of total,
extinction-corrected \bmv{} colour for Sample 2m ($D \leq 30$ Mpc,
$\logmstar \ge 9$). These should be compared with the plots of bar
fraction versus \gmr{} colour in the middle panels of
Figure~\ref{fig:fbar-vs-all}.\label{fig:fbar-vs-BmV}}

\end{figure*}

To test whether the colour transformation from \bmv{} to \gmr{}
introduces any significant biases, I plot the bar fraction as a function
of \bmvtc{} in Figure~\ref{fig:fbar-vs-BmV}. Comparison with the bar
fraction as a function of \gmr{} (middle panels of
Figure~\ref{fig:fbar-vs-all}) shows that the basic relation is
unchanged: bar fraction is approximately constant across a broad range
of colour, with some weak evidence for a decrease in the bar fraction
for redder galaxies.

\section{Comparison of Results for Alternate Subsamples}\label{sec:app-subsamples} 

\subsection{Subsample 2 versus Subsample 1}

\begin{figure*}
\begin{center}
\hspace*{-2mm}\includegraphics[scale=0.8]{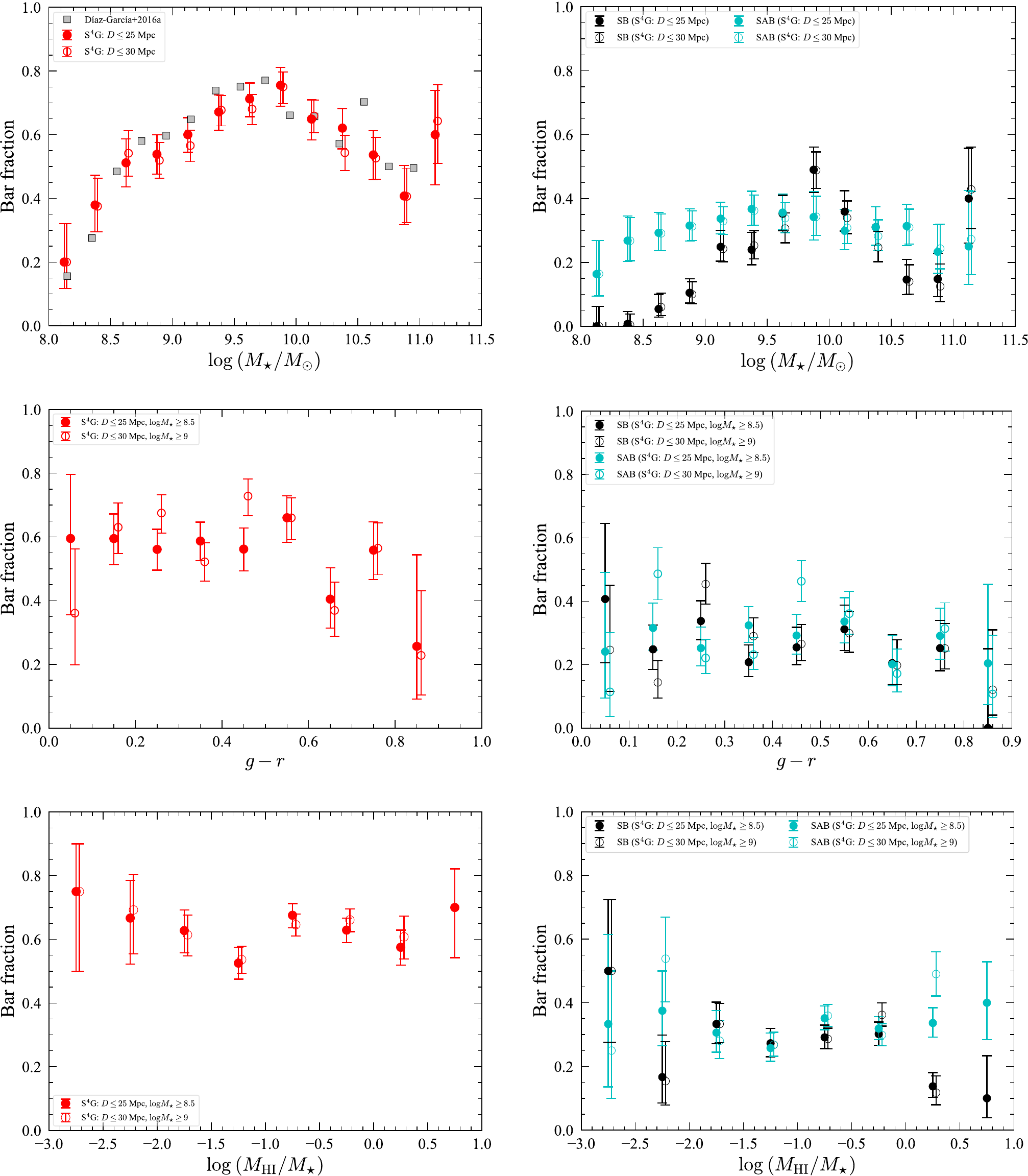}
\end{center}

\caption{As for Figure~\ref{fig:fbar-vs-all}, with filled circles
showing bar frequencies for Samples~1 (top row) or 1m (middle and bottom
rows), but also showing bar frequencies for Samples~2 or 2m using hollow
circles. Points for Samples 2 and 2m are slightly offset along the
horizontal axes to make comparisons easer. For clarity, bar frequencies
from other studies are not shown (see Figure~\ref{fig:fbar-vs-all});
however, the upper-left panel does include the published bar frequencies
for the parent sample as they appear in Fig.~19 of \citet{dg16a}.
Note that Sample~2m does not have enough galaxies in the highest-\fgas{}
bins for bar fractions to be calculated there (bottom panels).
\label{fig:fbar-vs-mstar-compare-samples}}

\end{figure*}

Most of the analysis in this paper uses \sfourg{} galaxies from Sample~1
($D \leq 25$ Mpc) or the mass-limited Sample~1m ($D \leq 25$ Mpc and
$\logmstar \geq 8.5$). In this section, I compare the effects of using
Samples~2 ($D \leq 30$ Mpc) and 2m ($D \leq 30$ Mpc and $\logmstar \geq
9$). Figure~\ref{fig:fbar-vs-mstar-compare-samples} is equivalent to
Figure~\ref{fig:fbar-vs-all}, except that bar frequencies using both
samples are plotted against \logmstar, \gmr, and \fgas; frequencies
based on Samples~2 or 2m are plotted using hollow circles. The overall
trends are the same in all cases, with significant differences between
the samples only in the bluest or reddest \gmr{} bins, where the number
of galaxies is very small. For comparison, the bar fractions
reported by \citet{dg16a} based on their complete sample -- the basis
for the Parent Disc Sample of this paper -- are plotted
using gray squares.

\subsection{Bar Fraction in S0 Galaxies}\label{app:S0s}

\begin{figure*}
\begin{center}
\hspace*{-15mm}\includegraphics[scale=0.55]{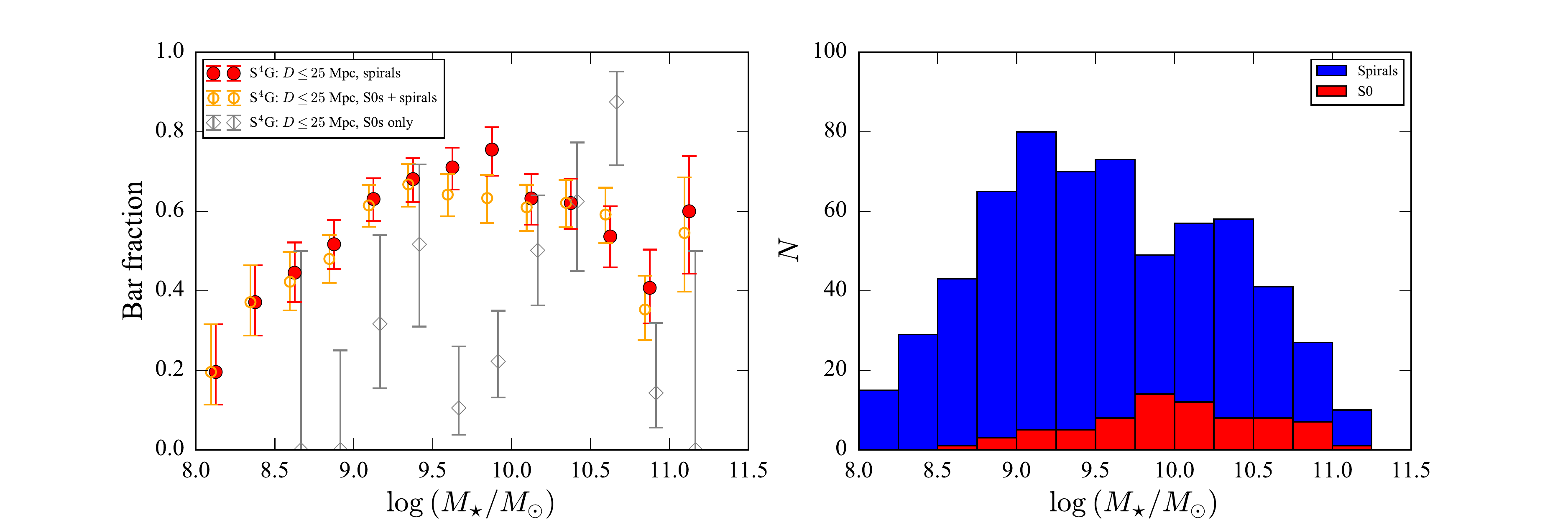}
\end{center}

\caption{Left: Bar fraction as a function of stellar mass for Sample~1
spirals (red circles, same as in Figure~\ref{fig:fbar-vs-all}), spirals +
S0 galaxies with same distance limit (open orange circles), and S0
galaxies alone (open grey diamonds). The flatter values of \fbar{}
between $\logmstar \sim 9.5$ and 10.5 for the spirals + S0s sample compared
to the pure-spirals sample is due primarily to the very low \fbar{}
values for S0 galaxies between $\logmstar \sim 9.5$ and 10 and
secondarily to the spike in S0 \fbar{} for $\logmstar \sim 10.5$. Right:
Distribution of stellar masses for spirals in Sample~1 (blue) and for
S0s meeting same criteria (red). \label{fig:fbar-spirals+S0-vs-mstar}}

\end{figure*}

Figure~\ref{fig:fbar-spirals+S0-vs-mstar} shows the effects of excluding
or including S0 galaxies in the sample, as well as the bar frequency (as
a function of stellar mass) for just the S0 galaxies. The red points
(left-hand panel) are the standard Sample~1 \fbar{} trend for spiral
galaxies, identical to the trend in the upper left panel of
Figure~\ref{fig:fbar-vs-all}. Adding the S0 galaxies to Sample~1
produces a flatter trend for $\logmstar > 9.5$ (orange points). The
right-hand panel compares the stellar-mass distributions of the spirals
and the S0s in the Parent Disc Sample. In contrast to the spirals, whose
mass distribution peaks at around $10^{9} \Msun$, the S0 distribution
peaks at $\sim 10^{10} \Msun$; this helps explain why including the S0s
mainly affects the higher-mass side of the \fbar{} trend in the
left-hand panel.

\bsp	
\label{lastpage}
\end{document}